\documentclass[12pt,a4paper]{article}
\usepackage{a4wide}
\usepackage{amsmath}
\usepackage{amssymb}
\usepackage{epsfig}
\usepackage{subfigure}
\usepackage{exscale}
\usepackage{float}
\usepackage{bbm}
\usepackage[numbers,sort&compress]{natbib}
\newcommand{\1}{{\mathbbm{1}}}
\newcommand{\p}{\partial}
\newcommand{\tild}{\widetilde}

\newcommand{\ontopof}[2]{\genfrac{}{}{0pt}{}{#1}{#2}}
\makeatletter
\@addtoreset{equation}{section}
\makeatother

\title{Two-Hole Bound States from a Systematic \\ 
Low-Energy Effective Field Theory for \\
Magnons and Holes in an Antiferromagnet}

\author{C.\ Br\"ugger$^a$, F.~K\"ampfer$^a$, M.~Moser$^a$,
  M.\ Pepe$^b$,\\ and U.-J.~Wiese$^a$
\\ \\
$^a$ Institute for Theoretical Physics, Bern University \\
Sidlerstrasse 5, CH-3012 Bern, Switzerland \\ \\
$^b$ Istituto Nazionale di Fisica Nucleare and \\
Dipartimento di Fisica, Universit\`a di Milano-Bicocca \\
3 Piazza della Scienza, 20126 Milano, Italy \\ \\}

\date{June 29, 2006}

\begin{document} 
\maketitle

\vspace{-1cm}

\begin{abstract} \normalsize

Identifying the correct low-energy effective theory for magnons and holes in an
antiferromagnet has remained an open problem for a long time. In analogy to the
effective theory for pions and nucleons in QCD, based on a symmetry analysis of
Hubbard and $t$-$J$-type models, we construct a systematic low-energy effective
field theory for magnons and holes located inside pockets centered at lattice 
momenta $(\pm \frac{\pi}{2a},\pm \frac{\pi}{2a})$. The effective theory is 
based on a nonlinear realization of the spontaneously broken spin symmetry and 
makes model-independent universal predictions for the entire class of lightly 
doped antiferromagnetic precursors of high-temperature superconductors. The 
predictions of the effective theory are exact, order by order in a systematic 
low-energy expansion. We derive the one-magnon exchange potentials between two 
holes in an otherwise undoped system. Remarkably, in some cases the
corresponding two-hole Schr\"odinger equations can even be solved analytically.
The resulting bound states have $d$-wave characteristics. The ground state wave
function of two holes residing in different hole pockets has a 
$d_{x^2-y^2}$-like symmetry, while for two holes in the same pocket the 
symmetry resembles $d_{xy}$.

\end{abstract}
 
\maketitle
 
\newpage

\section{Introduction}

The discovery of high-temperature superconductors \cite{Bed86} has motivated 
numerous studies of their doped antiferromagnetic precursors. In particular, 
the dynamics of holes in an antiferromagnet have been investigated in great 
detail in the condensed matter literature
\cite{Bri70,Hir85,And87,Gro87,Shr88,Tru88,Sch88,Kan89,Sac89,Wen89,Kra89,Sha90,And90,Sin90,Tru90,Els90,Dag90,Vig90,Kop90,Sch90,Wen91,Ver91,Mar91,Mon91,Liu92,Kue93,Dah93,Kuc93,Fla93,Sus94,Dag94,Bel95,Alt95,Kyu97,Chu98,Kar98,Ape98,Bru00,Mis01,Sac03,Sus04,Kot04}.
However, a systematic investigation of these dynamics is complicated due to the
strong correlations between the electrons in these systems. Unfortunately, away
from half-filling, the standard microscopic Hubbard and $t$-$J$-type models 
cannot be solved numerically due to a severe fermion sign problem. Analytic 
calculations, on the other hand, usually suffer from uncontrolled 
approximations. Substantial progress has been made in the pioneering work of 
Chakravarty, Halperin, and Nelson \cite{Cha89} who described the low-energy 
magnon physics by an effective field theory --- the
$(2+1)$-d $O(3)$-invariant nonlinear $\sigma$-model. Based on this work,
starting with Wen \cite{Wen89} and Shankar \cite{Sha90}, there have been a 
number of approaches \cite{Sch90,Wen91,Kue93} that address the physics of both
magnons and holes using effective field theories. All these approaches use
composite vector fields to couple magnons and holes. The spin then appears as
the ``charge'' of an Abelian gauge field. In this context, confinement of the 
spin ``charge'' and resulting spin-charge separation has sometimes been 
invoked. In these approaches an effective Lagrangian is usually obtained from 
an underlying microscopic system (e.g.\ from the Hubbard or $t$-$J$ model) by 
integrating out high-energy degrees of freedom. In this manner a variety of 
effective theories has been constructed. Unfortunately, there seems to be no 
agreement even on basic issues like the fermion field content of the effective 
theory or on the question how various symmetries are realized on those fields. 
In particular, it has never been demonstrated convincingly that any of the 
effective theories proposed so far indeed correctly describes the low-energy 
physics of the underlying microscopic systems quantitatively. 

The experience with chiral perturbation theory for the strong interactions 
shows that effective field theory is able to provide a systematic --- i.e.\ 
order by order exact --- description of the low-energy physics of 
nonperturbative systems as complicated as QCD. One main goal of this paper is 
to provide the same for the antiferromagnetic precursors of high-temperature 
superconductors. Inspired by strongly interacting systems in particle physics 
\cite{Wei79,Gas85}, we have recently approached the problem of constructing a 
low-energy effective field theory for magnons and holes in a systematic 
manner \cite{Kae05}. A central ingredient is the nonlinear realization of the 
spontaneously broken global symmetry \cite{Col69,Cal69} --- in this case of the
$SU(2)_s$ spin symmetry. This again leads to the same composite vector fields
that appeared in previous approaches to the problem. In particular, spin again
appears as an Abelian ``charge'' to which a composite magnon ``gauge'' field
couples. However, this gauge field does not mediate confining interactions. It
just mediates magnon exchange, which represents a weak interaction at low 
energies. Consequently, spin-charge separation does not arise. In analogy to 
baryon chiral perturbation theory \cite{Geo84,Gas88,Jen91,Ber92,Bec99} --- the 
effective theory for pions and nucleons --- we have extended the pure magnon 
effective theory of 
\cite{Cha89,Neu89,Fis89,Has90,Has91,Has93,Chu94,Leu94,Hof99,Rom99,Bae04} by 
including charge carriers. In \cite{Kae05} we have investigated the simplest
case of charge carriers appearing at lattice momenta $(0,0)$ or 
$(\frac{\pi}{a},\frac{\pi}{a})$ in the Brillouin zone. However, angle resolved 
photoemission spectroscopy (ARPES) experiments \cite{Wel95,LaR97,Kim98,Ron98} 
as well as theoretical investigations \cite{Tru88,Shr88,Els90,Bru00,Mis01} show
that doped holes appear inside hole pockets centered at the lattice momenta 
$(\pm \frac{\pi}{2a},\pm \frac{\pi}{2a})$. In this paper, we generalize the 
effective theory of \cite{Kae05} to this case. 

It should be pointed out that the effective theory to be constructed below is 
based on microscopic systems such as the Hubbard or $t$-$J$ model, but does not
necessarily reflect all aspects of the actual cuprate materials. For example, 
just like the Hubbard or $t$-$J$ model, the effective theory does not contain
impurities which are a necessary consequence of doping in the real materials. 
Also long-range Coulomb forces, anisotropies, or the effects of small couplings
between different $CuO_2$ layers are neglected in the effective theory. 
Furthermore, the underlying crystal lattice is imposed as a rigid structure by 
hand, such that phonons are excluded from the outset. Although all these 
effects can in principle be incorporated in the effective theory, for the 
moment we exclude them, in order not to obscure the basic physics of magnons 
and holes. As a consequence, the effective theory does not describe the actual 
materials in all details. Still, it should be pointed out that the predictions 
of the effective theory are not limited to just the Hubbard or $t$-$J$ model, 
but are universally applicable to a wide range of microscopic systems. In fact,
the low-energy physics of any antiferromagnet that possesses the assumed 
symmetries and has hole pockets at $(\pm \frac{\pi}{2a},\pm \frac{\pi}{2a})$ is
described correctly, order by order in a systematic low-energy expansion. 
Material-specific properties enter the effective theory in the form of a priori
undetermined low-energy parameters, such as the spin stiffness or the spinwave
velocity. The values of the low-energy parameters for a concrete underlying
microscopic system can be determined by comparison with experiments or with
numerical simulations. For example, precise numerical simulations of low-energy
observables in the $t$-$J$ model constitute a most stringent test of the 
effective theory. Such simulations are presently in progress.

After constructing the effective theory, we use it to calculate 
the one-magnon exchange potentials between two holes and we solve the 
corresponding Schr\"odinger equations. Remarkably, in some cases the 
Schr\"odinger equations can be solved completely analytically. The location of 
the hole pockets has an important effect on the dynamics and implies $d$-wave 
characteristics of hole pairs. Using the methods described in this paper, 
analogous to applications of baryon chiral perturbation theory to few-nucleon 
systems \cite{Wei90,Kap98,Epe98,Bed98,Kol99,Par99,Epe01,Bea02,Bed02,Nog05}, we 
have recently investigated magnon-mediated binding between two holes residing 
in two different hole pockets \cite{Bru05}. Here we discuss these issues in
more detail and we extend the investigation to a pair of holes in the same 
pocket. In this paper, we limit ourselves to an isolated pair of holes in an
otherwise undoped antiferromagnet. Lightly doped antiferromagnets will be 
investigated in a forthcoming publication \cite{Bru06}.

The paper is organized as follows. In section 2 the symmetries of the
microscopic Hubbard and $t$-$J$ models are summarized. Section 3 describes the
nonlinear realization of the spontaneously broken $SU(2)_s$ spin symmetry. In
section 4 the transformations of the effective fields for charge carriers are
related to the ones of the underlying microscopic models. The hole fields are 
identified and the electron fields are eliminated in section 5. Also the 
leading terms in the effective Lagrangian for magnons and holes are constructed
and accidental emergent flavor and Galilean boost symmetries are discussed. The
resulting one-magnon exchange potentials are derived and the corresponding 
Schr\"odinger equations are studied in section 6. Finally, section 7 contains 
our conclusions.

\section{Symmetries of Microscopic Models}

The standard microscopic models for antiferromagnetism and high-temperature 
superconductivity are Hubbard and $t$-$J$-type models. The symmetries of these 
models are of central importance for the construction of the low-energy 
effective theories for magnons and charge carriers. The Hubbard model is 
defined by the Hamiltonian
\begin{equation}
H = - t \sum_{x, i} (c_x^\dagger c_{x+\hat i} + 
c_{x + \hat i}^\dagger c_x) + 
\frac{U}{2} \sum_x (c_x^\dagger c_x - 1)^2 - \mu \sum_x (c_x^\dagger c_x - 1).
\end{equation}
Here $x$ denotes the sites of a 2-dimensional square lattice and $\hat i$ is a 
vector of length $a$ (where $a$ is the lattice spacing) pointing in the 
$i$-direction. Furthermore, $t$ is the nearest-neighbor hopping parameter, 
while $U>0$ is the strength of the screened on-site Coulomb repulsion, and 
$\mu$ is the chemical potential for fermion number relative to half-filling. 
The fermion creation and annihilation operators are given by
\begin{equation}
c_x^\dagger = (c_{x \uparrow}^\dagger, c_{x \downarrow}^\dagger), \qquad
c_x = \left(\begin{array}{c} c_{x \uparrow} \\ c_{x \downarrow} 
\end{array} \right).
\end{equation}
They obey standard anticommutation relations. The $SU(2)_s$ symmetry is 
generated by the total spin
\begin{equation}
\label{spin}
\vec S = \sum_x \vec S_x = \sum_x c_x^\dagger \ \frac{\vec \sigma}{2} \ c_x,
\end{equation}
where $\vec \sigma$ are the Pauli matrices, while the $U(1)_Q$ fermion number
(relative to half-filling) is generated by the charge operator 
\begin{equation}
Q = \sum_x Q_x = \sum_x (c_x^\dagger c_x - 1)
  = \sum_x (c_{x \uparrow}^\dagger c_{x \uparrow} +
    c_{x \downarrow}^\dagger c_{x \downarrow} - 1) .
\end{equation}
For $\mu = 0$, the Hubbard model even possesses a non-Abelian $SU(2)_Q$ 
extension of the fermion number symmetry \cite{Zha90,Yan90} which is generated 
by 
\begin{equation}
Q^+ = \sum_x (-1)^x c_{x \uparrow}^\dagger c_{x \downarrow}^\dagger, \quad
Q^- = \sum_x (-1)^x c_{x \downarrow} c_{x \uparrow}, \quad
Q^3 = \frac{1}{2} Q.
\end{equation}
The factor $(-1)^x = (-1)^{(x_1+x_2)/a}$ distinguishes between the sites of the
even and odd sublattice. The points on the even sublattice have $(-1)^x = 1$ 
while the points on the odd sublattice have $(-1)^x = - 1$. As discussed in 
detail in \cite{Kae05}, it is useful to introduce a matrix-valued fermion 
operator
\begin{equation}
\label{Coperator}
C_x = \left(\begin{array}{cc} c_{x \uparrow} &
(-1)^x \ c^\dagger_{x \downarrow} \\ c_{x \downarrow} &
- (-1)^x c_{x \uparrow}^\dagger \end{array} \right),
\end{equation}
which displays both the $SU(2)_s$ and the $SU(2)_Q$ symmetries in a compact 
form. Under combined transformations $g \in SU(2)_s$ and $\Omega \in SU(2)_Q$ 
it transforms as
\begin{equation}
\label{trafog}
^{\vec Q}C_x' = g C_x \Omega^T.
\end{equation}
Due to the antiferromagnetic order near half-filling, another important 
symmetry is the displacement $D_i$ by one lattice spacing in the $i$-direction 
which acts as
\begin{equation}
^{D_i}C_x = C_{x+\hat i} \sigma_3.
\end{equation}
The appearance of $\sigma_3$ on the right is due to the factor $(-1)^x$. As
discussed in \cite{Kae05}, it is also useful to introduce a combination $D'_i$
of the displacement symmetry $D_i$ with the $SU(2)_s$ transformation 
$g=i \sigma_2$
\begin{equation}
^{D'_i}C_x = (i \sigma_2) C_{x+\hat i} \sigma_3.
\end{equation}
The Hamiltonian can then be expressed in the manifestly $SU(2)_s$-, $SU(2)_Q$-,
$D_i$-, and thus also $D'_i$-invariant form
\begin{equation}
\label{HF}
H = - \frac{t}{2} \sum_{x, i} \mbox{Tr}[C_x^\dagger C_{x+\hat i} +
C_{x+\hat i}^\dagger C_x] + 
\frac{U}{12} \sum_x \mbox{Tr}[C_x^\dagger C_x C_x^\dagger C_x] -
\frac{\mu}{2} \sum_x \mbox{Tr}[C_x^\dagger C_x \sigma_3].
\end{equation}
The chemical potential term is only $U(1)_Q$-invariant, while the other two 
terms are manifestly $SU(2)_Q$-invariant.

We also need to consider the 90 degrees rotation $O$ of the quadratic lattice. 
It acts on a point $x = (x_1,x_2)$ as $Ox = (- x_2,x_1)$. Under this symmetry 
the fermion operator matrix transforms as 
\begin{equation}
^OC_x = C_{Ox}. 
\end{equation}
Under the spatial reflection $R$ at the $x_1$-axis, which turns $x$ into 
$Rx = (x_1,- x_2)$, one obtains
\begin{equation}
^RC_x = C_{Rx}.
\end{equation} 
It should be noted that, due to the presence of the lattice, the Hubbard model 
is not invariant under Galilean boosts.

The $t$-$J$ model is defined by the Hamilton operator
\begin{equation}
H = P \bigg\{- t \sum_{x,i} 
(c_x^\dagger c_{x+\hat i} + c_{x+\hat i}^\dagger c_x) +
J \sum_{x,i} \vec S_x \cdot \vec S_{x+\hat i} - \mu \sum_x Q_x\bigg\} P.
\end{equation}
Now the operators act in a restricted Hilbert space of empty or at most singly 
occupied sites, while states with doubly occupied sites are eliminated from the
Hilbert space by the projection operator $P$. Hence, the $t$-$J$ model can only
be doped with holes but not with electrons. The $t$-$J$ model has the same 
symmetries as the Hubbard model, except that the $SU(2)_Q$ extension of the
$U(1)_Q$ fermion number symmetry, which relates electrons to holes in the 
Hubbard model, is now absent.

In the $t$-$J$ model, a single hole has been simulated rather accurately in 
\cite{Bru00,Mis01}.
Using a worm-cluster algorithm similar to the algorithm used in \cite{Mis01}, 
we have computed the single-hole dispersion relation shown in figure 1. The
energy $E(\vec p)$ of a hole is minimal when its lattice momentum 
$\vec p = (p_1,p_2)$ is located in a hole pocket centered at 
$(\pm \frac{\pi}{2a},\pm \frac{\pi}{2a})$.
\begin{figure}[t]
\begin{center}
\vspace{-2.4cm}
\epsfig{file=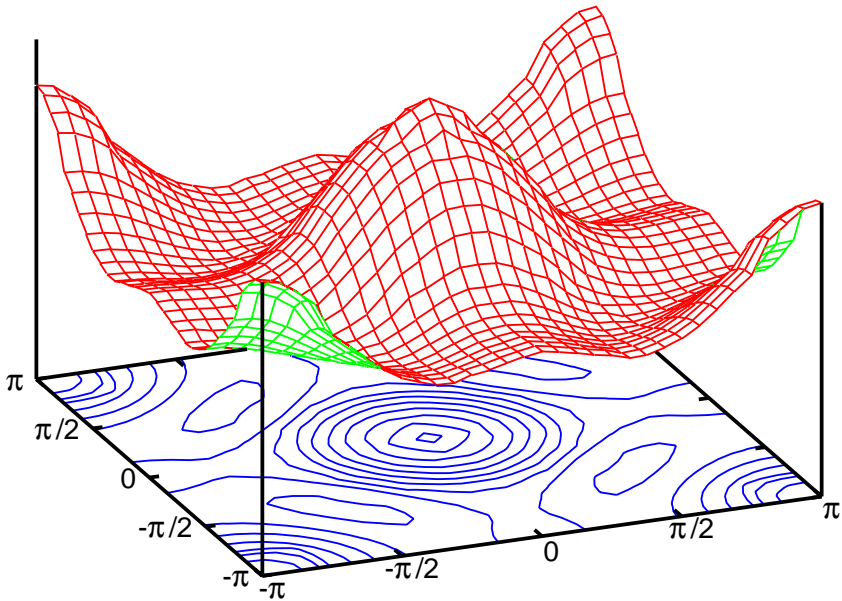,width=15cm}
\end{center}
\vspace{-1cm}
\caption{\it The dispersion relation $E(\vec p)$ of a single hole in the 
$t$-$J$ model (on a $32 \times 32$ lattice for $J = 2 t$) with hole pockets 
centered at $(\pm \frac{\pi}{2a},\pm \frac{\pi}{2a})$.}
\end{figure}

\section{Nonlinear Realization of the $SU(2)_s$ Symmetry}

The key to the low-energy physics of lightly doped cuprates is the spontaneous 
breakdown of the $SU(2)_s$ symmetry down to $U(1)_s$ which gives rise to two 
massless Nambu-Goldstone bosons --- the antiferromagnetic magnons. Analogous to
chiral perturbation theory for the Nambu-Goldstone pions in QCD \cite{Gas85}, a
systematic low-energy effective theory for magnons has been developed in 
\cite{Neu89,Fis89,Has90,Has91,Has93,Leu94,Hof99,Chu94,Rom99}. In order to 
couple charge carriers to the magnons, a nonlinear realization of the $SU(2)_s$
symmetry has been constructed in \cite{Kae05}. The global $SU(2)_s$ symmetry 
then manifests itself as a local $U(1)_s$ symmetry in the unbroken subgroup. 
This is analogous to baryon chiral perturbation theory in which the 
spontaneously broken $SU(2)_L \otimes SU(2)_R$ chiral symmetry of QCD is 
implemented on the nucleon fields as a local $SU(2)_{L=R}$ transformation in 
the unbroken isospin subgroup.

The staggered magnetization of an antiferromagnet is described by a unit-vector
field
\begin{equation}
\vec e(x) = (e_1(x),e_2(x),e_3(x)), \qquad \vec e(x)^2 = 1,
\end{equation}
in the coset space $SU(2)_s/U(1)_s = S^2$. Here $x = (x_1,x_2,t)$ denotes
a point in space-time. An equivalent $CP(1)$ representation uses $2 \times 2$ 
Hermitean projection matrices $P(x)$ that obey
\begin{equation}
P(x) = \frac{1}{2}(\1 + \vec e(x) \cdot \vec \sigma), \quad
P(x)^\dagger = P(x), \quad \mbox{Tr} P(x) = 1, \quad P(x)^2 = P(x).
\end{equation}
To leading order, the Euclidean magnon effective action takes the form
\begin{equation}
\label{Paction}
S[P] = \int d^2x \ dt \ \rho_s \mbox{Tr}\Big[\p_i P \p_i P +
\frac{1}{c^2} \p_t P \p_t P\Big].
\end{equation}
The index $i \in \{1,2\}$ labels the two spatial directions, while the index 
$t$ refers to the time-direction. The parameter $\rho_s$ is the spin stiffness 
and $c$ is the spinwave velocity. As discussed in detail in \cite{Kae05}, the 
action is invariant under the symmetries of the corresponding microscopic 
models which are realized as follows in the effective theory
\begin{alignat}{3}
SU(2)_s:&\quad &P(x)' &= g P(x) g^\dagger, \nonumber \\
SU(2)_Q:&\quad  &^{\vec Q}P(x) &= P(x), \nonumber \\
D_i:&\quad &^{D_i}P(x) &= \1 - P(x), \nonumber \\
D'_i:&\quad &^{D'_i}P(x) &= P(x)^*, \nonumber \\
O:&\quad &^OP(x) &= P(Ox), &\quad Ox &= (- x_2,x_1,t), \nonumber \\
R:&\quad &^RP(x) &= P(Rx), &\quad Rx &= (x_1,- x_2,t), \nonumber \\
T:&\quad &^TP(x) &= \1 - P(Tx), &\quad Tx &= (x_1,x_2,- t), \nonumber \\
T':&\quad &^{T'}P(x) &= (i \sigma_2) \; \left[^TP(x)\right] 
(i \sigma_2)^\dagger   = P(Tx)^*. \hspace{-4em}
\end{alignat}
Here $T$ denotes time-reversal. The symmetry $T'$ combines $T$ with the
$SU(2)_s$ rotation $g = i \sigma_2$.

The definition of the nonlinear realization of the $SU(2)_s$ symmetry proceeds
as follows. First, one diagonalizes the magnon field by a unitary 
transformation 
$u(x) \in SU(2)$, i.e.
\begin{equation}
u(x) P(x) u(x)^\dagger = \frac{1}{2}(\1 + \sigma_3) = 
\left(\begin{array}{cc} 1 & 0 \\ 0 & 0 \end{array} \right), \qquad 
u_{11}(x) \geq 0.
\end{equation}
Parameterizing
\begin{equation}
\label{evec}
\vec e(x) = 
(\sin\theta(x) \cos\varphi(x),\sin\theta(x) \sin\varphi(x),\cos\theta(x)),
\end{equation}
one obtains
\begin{equation}
u(x) = \left(\begin{array}{cc} \cos(\frac{1}{2}\theta(x)) & 
\sin(\frac{1}{2}\theta(x)) \exp(- i \varphi(x)) \\
- \sin(\frac{1}{2}\theta(x)) \exp(i \varphi(x)) & \cos(\frac{1}{2}\theta(x)) 
\end{array}\right).
\end{equation}
Under a global $SU(2)_s$ transformation $g$, the diagonalizing field $u(x)$
transforms as
\begin{equation}
\label{trafou}
u(x)' = h(x) u(x) g^\dagger, \qquad u_{11}(x)' \geq 0,
\end{equation}
which implicitly defines the nonlinear symmetry transformation 
\begin{equation}
h(x) = \exp(i \alpha(x) \sigma_3)
  = \left(\begin{array}{cc}
  \exp(i \alpha(x)) & 0 \\ 0 & \exp(- i \alpha(x)) \end{array} \right)
  \in U(1)_s.
\end{equation}
Under the displacement symmetry $D_i$ the staggered magnetization changes 
sign, i.e.\ $^{D_i}\vec e(x) = -\vec e(x)$, such that one obtains 
\begin{equation}
^{D_i}u(x) = \tau(x) u(x),
\end{equation} 
with
\begin{equation}
\label{taueq}
\tau(x) = \left(\begin{array}{cc} 0 & - \exp(- i \varphi(x)) \\
\exp(i \varphi(x)) & 0 \end{array} \right).
\end{equation}
Introducing the traceless anti-Hermitean field
\begin{equation}
v_\mu(x) = u(x) \p_\mu u(x)^\dagger,
\end{equation}
one obtains the following transformation rules
\begin{alignat}{3}
SU(2)_s:&\quad &v_\mu(x)' &= h(x) [v_\mu(x) + \p_\mu] h(x)^\dagger,
  \hspace{-5em} \nonumber \\
SU(2)_Q:&\quad &^{\vec Q}v_\mu(x) &= v_\mu(x), \nonumber \\
D_i:&\quad &^{D_i}v_\mu(x) &= \tau(x)[v_\mu(x) + \p_\mu] \tau(x)^\dagger,
  \hspace{-5em} \nonumber \\
D'_i:&\quad &^{D'_i}v_\mu(x) &= v_\mu(x)^*, \nonumber \\
O:&\quad &^Ov_i(x) &= \varepsilon_{ij} v_j(Ox), \quad
  &^Ov_t(x) &= v_t(Ox), \nonumber \\
R:&\quad &^Rv_1(x) &= v_1(Rx), \quad &^Rv_2(x) &= - v_2(Rx),
  \quad ^Rv_t(x) = v_t(Rx), \nonumber \\
T:&\quad &^Tv_j(x) &= \ ^{D_i}v_j(Tx), \quad &^Tv_t(x) &= - \ ^{D_i}v_t(Tx), 
  \nonumber \\
T':&\quad &^{T'}v_j(x) &= \ ^{D'_i}v_j(Tx), \quad
   &^{T'}v_t(x) &= - ^{D'_i}v_t(Tx).
\end{alignat}
Writing
\begin{equation}
v_\mu(x) = i v_\mu^a(x) \sigma_a, \qquad
v_\mu^\pm(x) = v_\mu^1(x) \mp i v_\mu^2(x),
\end{equation}
the field $v_\mu(x)$ decomposes into an Abelian ``gauge'' field $v_\mu^3(x)$ 
and two ``charged'' vector fields $v_\mu^\pm(x)$.

\section{Transformation Rules of Charge Carrier Fields}

Due to the nonperturbative dynamics, it is impossible in practice to rigorously
derive the low-energy effective theory from the underlying microscopic physics.
Still, in this section we will attempt to relate the transformation rules of 
the effective fields describing the charge carriers to those of the microscopic
fermion operator matrix $C_x$ of eq.(\ref{Coperator}).

\subsection{Sublattice Fermion Fields}

In \cite{Kae05} we have introduced operators $\Psi^A_x$ and $\Psi^B_x$ on the 
even and odd sublattices $\Psi^{A,B}_x = u(x) C_x$ with $(-1)^x = 1$ for $A$
and $(-1)^x = - 1$ for $B$. In the 
Brillouin zone the corresponding linear combinations $\Psi^A_x \pm \Psi^B_x$ 
are located at lattice momenta $(0,0)$ and $(\frac{\pi}{a},\frac{\pi}{a})$. In 
order to account for the experimentally observed \cite{Wel95,LaR97,Kim98,Ron98}
as well as theoretically predicted \cite{Tru88,Shr88,Els90,Bru00,Mis01} hole 
pockets centered at $(\pm \frac{\pi}{2a},\pm \frac{\pi}{2a})$, we now introduce
eight sublattices $A$, $B$, ..., $H$ as illustrated in figure 2.
\begin{figure}[t]
\begin{center}
\vspace{-0.4cm}
\epsfig{file=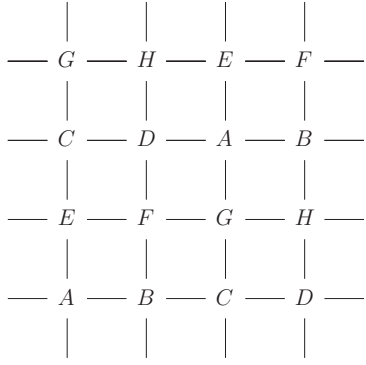,width=5cm}
\end{center}
\caption{\it The layout of the eight sublattices $A$, $B$, ..., $H$.}
\end{figure}
While it would be unnatural to introduce more than two sublattices in a pure 
antiferromagnet, the eight sublattices are a natural and even necessary concept
when one wants to describe fermions located in hole pockets centered at
$(\pm \frac{\pi}{2a},\pm \frac{\pi}{2a})$. We now introduce new lattice 
operators 
\begin{equation}
\Psi^{A,B,...,H}_x = u(x) C_x, \qquad x \in A, B, ... , H,
\end{equation}
which inherit their transformation properties from the operators of the Hubbard
model. According to eqs.(\ref{trafou}) and (\ref{trafog}), under the $SU(2)_s$ 
symmetry one obtains
\begin{equation}
{\Psi_x^X}' = u(x)' C_x' = h(x) u(x) g^\dagger g C_x = 
h(x) \Psi_x^X, \qquad X \in \{A,B,...,H\}. 
\end{equation}
Similarly, under the $SU(2)_Q$ symmetry one obtains
\begin{equation}
^{\vec Q}\Psi_x^X = {}^{\vec Q}u(x) ^{\vec Q}C_x = u(x) C_x \Omega^T 
= \Psi_x^X \Omega^T. 
\end{equation}
Under the displacement symmetry the new operators transform as
\begin{equation}
^{D_i}\Psi^X_x = {}^{D_i}u(x+\hat i) C_{x+\hat i} \sigma_3 = 
\tau(x+\hat i) u(x+\hat i) C_{x+\hat i} \sigma_3 = 
\tau(x+\hat i) \Psi^{D_iX}_{x+\hat i} \sigma_3,
\end{equation}
where $\tau(x)$ is the field introduced in eq.(\ref{taueq}) and $D_iX$ is the
sublattice that one obtains by shifting sublattice $X$ by one lattice spacing
in the $i$-direction. Similarly, under the symmetry $D'_i$ one finds
\begin{equation}
^{D'_i}\Psi^X_x = {}^{D'_i}u(x+\hat i) (i \sigma_2) C_{x+\hat i} \sigma_3 = 
u(x+\hat i)^* (i \sigma_2) C_{x+\hat i} \sigma_3 = 
(i \sigma_2) \Psi^{D_iX}_{x+\hat i} \sigma_3,
\end{equation}
while under the 90 degrees rotation $O$
\begin{equation}
^O\Psi^X_x = {}^Ou(x) \, ^OC_x = u(Ox) C_{Ox} = \Psi^{OX}_{Ox},
\end{equation}
and under the reflection $R$
\begin{equation}
^R\Psi^X_x = {}^Ru(x) \, ^RC_x = u(Rx) C_{Rx} = \Psi^{RX}_{Rx}.
\end{equation}
Here $OX$ and $RX$ are the sublattices obtained by rotating or reflecting the
sublattice $X$. We arbitrarily chose the origin to lie on sublattice $A$.

In the low-energy effective theory we will use a path integral description 
instead of the Hamiltonian description used in the Hubbard model. In the 
effective theory the electron and hole fields are thus represented by 
independent Grassmann numbers $\psi^{A,B,...,H}_\pm(x)$ and 
$\psi^{A,B,...,H\dagger}_\pm(x)$ which are combined to
\begin{eqnarray}
\label{phi}
&&\Psi^X(x) = \left(\begin{array}{cc} \psi^X_+(x) & \psi^{X\dagger}_-(x) \\ 
\psi^X_-(x) & - \psi^{X\dagger}_+(x) \end{array} \right), \qquad
  X \in \{A,C,F,H\},
\nonumber \\
&&\Psi^X(x) = \left(\begin{array}{cc} \psi^X_+(x) & 
- \psi^{X\dagger}_-(x) \\ 
\psi^X_-(x) & \psi^{X\dagger}_+(x) \end{array} \right), \qquad
  X \in \{B,D,E,G\}.
\end{eqnarray}
Note that the even sublattices $A, C, F, H$ (with $(-1)^x = 1$) are treated
differently than the odd sublattices $B, D, E, G$ (with $(-1)^x = - 1$).
For notational convenience, we also introduce the fields
\begin{eqnarray}
\label{phidagger}
&&\Psi^{X\dagger}(x) = \left(\begin{array}{cc} 
\psi^{X\dagger}_+(x) & \psi^{X\dagger}_-(x) \\ \psi^X_-(x) & - \psi^X_+(x) 
\end{array} \right), \qquad X \in \{A,C,F,H\}, \nonumber \\
&&\Psi^{X\dagger}(x) = \left(\begin{array}{cc}
\psi^{X\dagger}_+(x) & \psi^{X\dagger}_-(x) \\ - \psi^X_-(x) & \psi^X_+(x)
\end{array} \right), \qquad X \in \{B,D,E,G\},
\end{eqnarray}
which consist of the same Grassmann fields $\psi^X_\pm(x)$ and 
$\psi^{X\dagger}_\pm(x)$ as $\Psi^X(x)$.

In contrast to the lattice operators, the fields $\Psi^X(x)$ are 
defined in the continuum. Hence, under the displacement symmetries $D_i$ and 
$D'_i$ we no longer distinguish between the points $x$ and $x+\hat i$. As a 
result, the transformation rules of the various symmetries take the form
\begin{alignat}{2}
SU(2)_s:&\quad &\Psi^X(x)' &= h(x) \Psi^X(x), \nonumber \\
SU(2)_Q:&\quad &^{\vec Q}\Psi^X(x) &= \Psi^X(x) \Omega^T, \nonumber \\
D_i:&\quad &^{D_i}\Psi^X(x) &= \tau(x) \Psi^{D_iX}(x) \sigma_3, \nonumber \\
D'_i:&\quad &^{D'_i}\Psi^X(x) &= (i \sigma_2) \Psi^{D_iX}(x) \sigma_3, 
\nonumber \\
O:&\quad &^O\Psi^X(x) &= \Psi^{OX}(Ox), \nonumber \\
R:&\quad &^R\Psi^X(x) &= \Psi^{RX}(Rx), \nonumber \\
T:&\quad &^T\Psi^X(x) &= 
\tau(Tx) (i \sigma_2) \left[\Psi^{X\dagger}(Tx)^T\right] \sigma_3, \nonumber \\
  &\quad &^T\Psi^{X\dagger}(x) &=
 - \sigma_3 \left[\Psi^X(Tx)^T\right] (i \sigma_2)^\dagger \tau(Tx)^\dagger, 
\nonumber \\
T':&\quad &^{T'}\Psi^X(x) &= - \left[\Psi^{X\dagger}(Tx)^T\right] \sigma_3, 
\nonumber \\
   &\quad &^{T'}\Psi^{X\dagger}(x) &= \sigma_3 \left[\Psi^X(Tx)^T\right].
\end{alignat}
Note that an upper index $T$ on the right denotes transpose, while on the left 
it denotes time-reversal. The form of the time-reversal symmetry $T$ in the 
effective theory with nonlinearly realized $SU(2)_s$ symmetry follows from the
usual form of time-reversal in the path integral of a nonrelativistic theory in
which the spin symmetry is linearly realized. The fermion fields in the two 
formulations just differ by a factor $u(x)$. In components the transformation 
rules take the form
\begin{alignat}{2}
SU(2)_s:&\quad &\psi^X_\pm(x)' &= \exp(\pm i \alpha(x)) \psi^X_\pm(x),
\nonumber \\
U(1)_Q:&\quad &^Q\psi^X_\pm(x) &= \exp(i \omega) \psi^X_\pm(x),
\nonumber \\
D_i:&\quad &^{D_i}\psi^X_\pm(x) &= \mp \exp(\mp i \varphi(x)) 
\psi^{D_iX}_\mp(x), \nonumber \\
D'_i:&\quad &^{D'_i}\psi^X_\pm(x) &= \pm \psi^{D_iX}_\mp(x),
\nonumber \\
O:&\quad &^O\psi^X_\pm(x) &= \psi^{OX}_\pm(Ox), \nonumber \\
R:&\quad &^R\psi^X_\pm(x) &= \psi^{RX}_\pm(Rx), \nonumber \\
T:&\quad &^T\psi^X_\pm(x) &= \exp(\mp i \varphi(Tx)) \psi^{X\dagger}_\pm(Tx),
  \nonumber \\
  &\quad &^T\psi^{X\dagger}_\pm(x) &= - \exp(\pm i \varphi(Tx)) \psi^X_\pm(Tx),
  \nonumber \\
T':&\quad &^{T'}\psi^X_\pm(x) &= - \psi^{X\dagger}_\pm(Tx), \nonumber \\
   &\quad &^{T'}\psi^{X\dagger}_\pm(x) &= \psi^X_\pm(Tx).
\end{alignat}

\subsection{Fermion Fields in Momentum Space Pockets}

Instead of working with the eight sublattice indices $X \in \{A,B,...,H\}$, it 
is more convenient to introduce eight corresponding lattice momentum indices
\begin{equation}
k = (k_1,k_2) \in 
\left\{\big(0,0\big),\; \big(\frac{\pi}{a},\frac{\pi}{a}\big),\;
\big(\frac{\pi}{a},0\big),\; \big(0,\frac{\pi}{a}\big),\;
\big(\pm \frac{\pi}{2a},\pm \frac{\pi}{2a}\big)\right\}.
\end{equation}
The eight sublattices represent a minimal set that allows us to address the
lattice momenta $(\pm \frac{\pi}{2a},\pm \frac{\pi}{2a})$ which define the
centers of hole pockets in the cuprates. By introducing further sublattices it
would be straightforward to reach other lattice momenta as well. With the 
present construction we restrict ourselves to the momenta listed above and
illustrated in figure 3.
\begin{figure}[t]
\begin{center}
\vspace{-0.4cm}
\epsfig{file=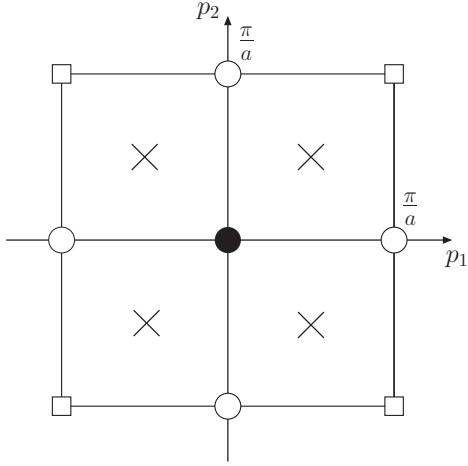,width=7cm}
\end{center}
\caption{\it The eight lattice momenta (and their periodic copies) dual to the 
eight sublattices $A$, $B$,...,$H$. In the cuprates the holes reside in 
momentum space pockets centered at lattice momenta 
$\big(\pm \frac{\pi}{2a},\pm \frac{\pi}{2a}\big)$ which are represented by the 
four crosses.}
\end{figure}
We now construct new fields
\begin{align}
\psi^k_\pm(x)=&\; \frac{1}{\sqrt{8}} \Big\{\psi^A_\pm(x) + 
e^{- i k_1 a} \psi^B_\pm(x) + e^{- 2 i k_1 a} \psi^C_\pm(x) + 
e^{- 3 i k_1 a} \psi^D_\pm(x) \nonumber \\
& + e^{- i k_2 a} [\psi^E_\pm(x) + e^{- i k_1 a} \psi^F_\pm(x) + 
e^{- 2 i k_1 a} \psi^G_\pm(x) + e^{- 3 i k_1 a} \psi^H_\pm(x)]\Big\}, 
\end{align}
which transform as
\begin{alignat}{2}
SU(2)_s:&\quad &\psi^k_\pm(x)' &= \exp(\pm i \alpha(x)) \psi^k_\pm(x),
\nonumber \\
U(1)_Q:&\quad &^Q\psi^k_\pm(x) &= \exp(i \omega) \psi^k_\pm(x),
\nonumber \\
D_i:&\quad &^{D_i}\psi^k_\pm(x) &= 
\mp \exp(i k_i a) \exp(\mp i \varphi(x)) \psi^k_\mp(x),
\nonumber \\
D'_i:&\quad &^{D'_i}\psi^k_\pm(x) &= \pm \exp(i k_i a) \psi^k_\mp(x),
\nonumber \\
O:&\quad &^O\psi^k_\pm(x) &= \psi^{Ok}_\pm(Ox), \nonumber \\
R:&\quad &^R\psi^k_\pm(x) &= \psi^{Rk}_\pm(Rx), \nonumber \\
T:&\quad &^T\psi^k_\pm(x) &= \exp(\mp i \varphi(Tx)) \psi^{-k\dagger}_\pm(Tx),
\nonumber \\
&\quad &^T\psi^{k\dagger}_\pm(x) &= - \exp(\pm i \varphi(Tx)) 
\psi^{-k}_\pm(Tx), \nonumber \\
T':&\quad &^{T'}\psi^k_\pm(x) &= - \psi^{-k\dagger}_\pm(Tx), \nonumber \\
&\quad &^{T'}\psi^{k\dagger}_\pm(x) &= \psi^{-k}_\pm(Tx).
\end{alignat}
Here $Ok$ and $Rk$ are the momenta obtained by rotating or reflecting the 
lattice momentum $k$. From the component fields one can again construct 
matrix-valued fields
\begin{equation}
\Psi^k(x) = \left(\begin{array}{cc} \psi^k_+(x) & \psi^{-k'\dagger}_-(x) 
\\ \psi^k_-(x) & - \psi^{-k' \dagger}_+(x) \end{array} \right), \quad
\Psi^{k\dagger}(x) = \left(\begin{array}{cc} 
\psi^{k\dagger}_+(x) & \psi^{k\dagger}_-(x) \\ 
\psi^{-k'}_-(x) & - \psi^{-k'}_+(x) \end{array} \right),
\end{equation}
with $k' = k + (\frac{\pi}{a},\frac{\pi}{a})$. The matrix-valued fields then 
transform as
\begin{alignat}{2}
\label{phitrafo}
SU(2)_s:&\quad &\Psi^k(x)' &= h(x) \Psi^k(x), \nonumber \\
SU(2)_Q:&\quad &^{\vec Q}\Psi^k(x) &= \Psi^k(x) \Omega^T, \nonumber \\
D_i:&\quad &^{D_i}\Psi^k(x) &= \exp(i k_i a) \tau(x) \Psi^k(x) \sigma_3, 
\nonumber \\
D'_i:&\quad &^{D'_i}\Psi^k(x) &= 
\exp(i k_i a) (i \sigma_2) \Psi^k(x) \sigma_3, \nonumber \\
O:&\quad &^O\Psi^k(x) &= \Psi^{Ok}(Ox), \nonumber \\
R:&\quad &^R\Psi^k(x) &= \Psi^{Rk}(Rx), \nonumber \\
T:&\quad &^T\Psi^k(x) &= 
\tau(Tx) (i \sigma_2) \left[\Psi^{-k\dagger}(Tx)^T\right] \sigma_3, 
\nonumber \\
&\quad &^T\Psi^{k\dagger}(x) &= - \sigma_3 \left[\Psi^{-k}(Tx)^T\right] 
(i \sigma_2)^\dagger \tau(Tx)^\dagger, \nonumber \\
T':&\quad &^{T'}\Psi^k(x) &= - \left[\Psi^{-k\dagger}(Tx)^T\right] \sigma_3, 
\nonumber \\
&\quad &^{T'}\Psi^{k\dagger}(x) &= \sigma_3 \left[\Psi^{-k}(Tx)^T\right].
\end{alignat}
In \cite{Kae05} we have limited ourselves to two sublattices $A$ and $B$ which 
leads to the lattice momenta $(0,0)$ and $(\frac{\pi}{a},\frac{\pi}{a})$. The 
main purpose of the present paper is to describe holes located in pockets 
centered at $(\pm \frac{\pi}{2a},\pm \frac{\pi}{2a})$.

\section{Effective Theory for Magnons and Holes}

From now on we will limit ourselves to theories with holes as the only charge 
carriers. In order to identify the hole and to eliminate the electron fields, 
we consider the most general mass terms consistent with the symmetries.

\subsection{Hole Field Identification and Electron Field Elimination}

It turns out that mass terms cannot mix the various lattice momenta 
arbitrarily. In particular, through a mass term a field $\psi^k_\pm(x)$ with 
lattice momentum $k$ can only mix with fields with lattice momenta $k$ or $k'$.
Hence, the eight lattice momenta can be divided into four pairs which are 
associated with three different cases. The simplest case in which $k = (0,0)$
and $k' = (\frac{\pi}{a},\frac{\pi}{a})$ has been investigated in great detail
in \cite{Kae05}, but is not realized in the cuprates. Another case in which
$k = (\frac{\pi}{a},0)$ and $k' = (0,\frac{\pi}{a})$ describes electron doping
and will be investigated elsewhere.  In the following, we concentrate on
hole-doped cuprates. In this case, the hole pockets are centered at lattice
momenta
\begin{align}
k^\alpha = \big(\frac{\pi}{2a}, \, \frac{\pi}{2a}\big), \quad
{k^\alpha}' = - k^\alpha, \qquad
k^\beta = \big(\frac{\pi}{2a}, - \frac{\pi}{2a}\big), \quad
{k^\beta}'  = - k^\beta.
\end{align}

Using the transformation rules of eq.(\ref{phitrafo}) one can construct the 
following invariant mass terms
\begin{align} 
\sum_{f=\alpha,\beta} & \frac{1}{2}
\mbox{Tr} \big[ {\cal M} (\Psi^{k^f\dagger} \sigma_3 \Psi^{{k^f}'} +
\Psi^{{k^f}'\dagger} \sigma_3 \Psi^{k^f}) +
m (\Psi^{k^f\dagger} \Psi^{k^f} \sigma_3 +
\Psi^{{k^f}'\dagger} \Psi^{{k^f}'} \sigma_3)
\big] \nonumber \\
= &\, \sum_{f=\alpha,\beta} \big[
{\cal M} \big(\psi^{k^f\dagger}_+ \psi^{{k^f}'}_+ - 
\psi^{k^f\dagger}_- \psi^{{k^f}'}_- +
\psi^{{k^f}'\dagger}_+ \psi^{k^f}_+ - 
\psi^{{k^f}'\dagger}_- \psi^{k^f}_- \big) 
\nonumber \\
&\,  \hspace{2.5em} + m \big(\psi^{k^f\dagger}_+ \psi^{k^f}_+ + 
\psi^{k^f\dagger}_- \psi^{k^f}_- + \psi^{{k^f}'\dagger}_+ \psi^{{k^f}'}_+ + 
\psi^{{k^f}'\dagger}_- \psi^{{k^f}'}_- \big) \big] \nonumber \\
= &\, \sum_{f=\alpha,\beta} \bigg[
\big(\psi^{k^f\dagger}_+, \, \psi^{{k^f}'\dagger}_+ \big) 
\bigg(\begin{array}{cc} m & {\cal M} \\ {\cal M} & m \end{array}\bigg)
\bigg(\begin{array}{c} \psi^{k^f}_+ \\ \psi^{{k^f}'}_+ \end{array}\bigg) 
\nonumber \\
&\, \hspace{2.5em} + \big(\psi^{k^f\dagger}_-, \, \psi^{{k^f}'\dagger}_- \big)
\bigg(\begin{array}{cc} m & - {\cal M} \\ - {\cal M} & m \end{array}\bigg)
\bigg(\begin{array}{c} \psi^{k^f}_- \\ \psi^{{k^f}'}_- \end{array}\bigg)
\bigg].
\end{align}
The terms proportional to ${\cal M}$ are $SU(2)_Q$-invariant while the terms
proportional to $m$ are only $U(1)_Q$-invariant. By diagonalizing the mass 
matrices one can identify particle and hole fields. The eigenvalues of the mass
matrices are $m \pm {\cal M}$. In the $SU(2)_Q$-symmetric case, i.e.\ for 
$m = 0$, there is a particle-hole symmetry. The particles correspond to 
positive energy states with eigenvalue ${\cal M}$ and the holes correspond to 
negative energy states with eigenvalue $- {\cal M}$. In the presence of 
$SU(2)_Q$-breaking terms these energies are shifted and particles now
correspond to states with eigenvalue $m + {\cal M}$, while holes correspond to 
states with eigenvalue $m - {\cal M}$. The hole fields are identified from the 
corresponding eigenvectors as
\begin{equation}
\psi^f_+(x) = \frac{1}{\sqrt{2}}
\big[ \psi^{k^f}_+(x) - \psi^{{k^f}'}_+(x) \big], \qquad
\psi^f_-(x) = \frac{1}{\sqrt{2}}
\big[ \psi^{k^f}_-(x) + \psi^{{k^f}'}_-(x) \big].
\end{equation}
It should be noted that processes involving electrons and holes simultaneously
cannot be treated in a systematic low-energy effective theory. Electrons and 
holes can annihilate, which turns their rest mass into other forms of energy.
This is necessarily a high-energy process. Only in the presence of an exact
$SU(2)_Q$ symmetry, the $SU(2)_Q$-nonsinglet electron-hole states are protected
against annihilation and can be treated systematically in a low-energy 
effective theory. Here we concentrate on the realistic case without $SU(2)_Q$ 
symmetry. Then electrons and holes must be considered separately. In this paper
we concentrate entirely on the holes. Under the various symmetries, the hole 
fields $\psi^f_\pm$ (with $f \in \{\alpha,\beta\}$) transform as
\begin{alignat}{2}
\label{symcomp}
SU(2)_s:&\quad &\psi^f_\pm(x)' &= \exp(\pm i \alpha(x)) \psi^f_\pm(x),
\nonumber \\
U(1)_Q:&\quad &^Q\psi^f_\pm(x) &= \exp(i \omega) \psi^f_\pm(x),
\nonumber \\
D_i:&\quad &^{D_i}\psi^f_\pm(x) &= 
\mp \exp(i k^f_i a) \exp(\mp i \varphi(x)) \psi^f_\mp(x),
\nonumber \\
D'_i:&\quad &^{D'_i}\psi^f_\pm(x) &= \pm \exp(i k^f_i a) \psi^f_\mp(x),
\nonumber \\
O:&\quad &^O\psi^\alpha_\pm(x) &= \mp \psi^\beta_\pm(Ox), \quad
^O\psi^\beta_\pm(x) = \psi^\alpha_\pm(Ox), \nonumber \\
R:&\quad &^R\psi^\alpha_\pm(x) &= \psi^\beta_\pm(Rx), \quad\;\;\;
^R\psi^\beta_\pm(x) = \psi^\alpha_\pm(Rx), \nonumber \\
T:&\quad &^T\psi^f_\pm(x) &= \pm \exp(\mp i \varphi(Tx)) 
\psi^{f\dagger}_\pm(Tx),
\nonumber \\
&\quad &^T\psi^{f\dagger}_\pm(x) &= \mp \exp(\pm i \varphi(Tx)) \psi^f_\pm(Tx),
\nonumber \\
T':&\quad &^{T'}\psi^f_\pm(x) &= \mp \psi^{f\dagger}_\pm(Tx), \nonumber \\
&\quad &^{T'}\psi^{f\dagger}_\pm(x) &= \pm \psi^f_\pm(Tx).
\end{alignat}
The action to be constructed below must be invariant under these symmetries.

\subsection{Effective Action for Magnons and Holes}

The terms in the action can be characterized by the (necessarily even) number 
$n_\psi$ of fermion fields they contain, i.e.
\begin{equation}
S[\psi^{f\dagger}_\pm,\psi^f_\pm,P] = \int d^2x \ dt \ \sum_{n_\psi}
{\cal L}_{n_\psi}
\end{equation}
The leading terms in the effective Lagrangian without fermion fields describe 
the pure magnon sector and take the form
\begin{equation}
{\cal L}_0 = \rho_s \mbox{Tr}
\big[ \p_i P \p_i P + \frac{1}{c^2} \p_t P \p_t P \big],
\end{equation}
with the spin stiffness $\rho_s$ and the spinwave velocity $c$. The leading
terms with two fermion fields (containing at most one temporal or two spatial
derivatives) describe the propagation of holes as well as their couplings to 
magnons and are given by
\begin{align}
{\cal L}_2=\sum_{\ontopof{f=\alpha,\beta}{\, s = +,-}} \Big[ &
M \psi^{f\dagger}_s \psi^f_s + \psi^{f\dagger}_s D_t \psi^f_s 
\nonumber \\[-3ex]
&+\frac{1}{2 M'} D_i \psi^{f\dagger}_s D_i \psi^f_s +
\sigma_f \frac{1}{2 M''} \big( D_1 \psi^{f\dagger}_s D_2 \psi^f_s +
D_2 \psi^{f\dagger}_s D_1 \psi^f_s \big) \nonumber \\[0.7ex]
&+\Lambda \big( \psi^{f\dagger}_s v^s_1 \psi^f_{-s} 
+ \sigma_f \psi^{f\dagger}_s v^s_2 \psi^f_{-s} \big) \nonumber \\
&+N_1 \psi^{f\dagger}_s v^s_i v^{-s}_i \psi^f_s +
\sigma_f N_2 \big( \psi^{f\dagger}_s v^s_1 v^{-s}_2 \psi^f_s + 
\psi^{f\dagger}_s v^s_2 v^{-s}_1 \psi^f_s \big) \Big].
\end{align}
Here $M$ is the rest mass and $M'$ and $M''$ are the kinetic masses of a hole, 
$\Lambda$ is a hole-one-magnon, and $N_1$ and $N_2$ are hole-two-magnon 
couplings, which all take real values. The sign $\sigma_f$ is $+$ for 
$f = \alpha$ and $-$ for $f = \beta$. The covariant derivatives are given by
\begin{eqnarray}
&&D_t \psi^f_\pm(x) = \left[\p_t \pm i v_t^3(x) - \mu \right] \psi^f_\pm(x),
\nonumber \\
&&D_i \psi^f_\pm(x) = \left[\p_i \pm i v_i^3(x)\right] \psi^f_\pm(x).
\end{eqnarray}
The chemical potential $\mu$ enters the covariant time-derivative like an
imaginary constant vector potential for the fermion number symmetry $U(1)_Q$.
As discussed in detail in \cite{Bae04,Kae05}, the coupling to external 
electromagnetic fields leads to further modifications of the covariant 
derivatives. Remarkably, the term in the action proportional to $\Lambda$ 
contains just a single (uncontracted) spatial derivative. Due to the nontrivial
rotation properties of flavor, this term is still 90 degrees rotation 
invariant. Due to the small number of derivatives it contains, this term 
dominates the low-energy dynamics. In particular, it alone is responsible for 
one-magnon exchange. Interestingly, although the effective theory of 
\cite{Kue93} has the same field content as the one presented here, the terms in
the Lagrangian are quite different. In particular, the term proportional to 
$\Lambda$ is absent in that theory and the physics is thus very different. This
also means that in \cite{Kue93} the symmetries are realized on the fermion 
fields in a different way. 

The above Lagrangian leads to a single hole dispersion relation
\begin{equation}
\label{dispersion}
E^f(\vec p) = M + \frac{p_i^2}{2 M'} + \sigma_f  \frac{p_1 p_2}{M''}.
\end{equation}
For $1/M'' = 0$ this would be the usual dispersion relation of a
free nonrelativistic particle. In that case, the hole pockets centered at
$(\frac{\pi}{2a},\pm \frac{\pi}{2a})$ would have a circular shape. However,
the 90 degrees rotation symmetry of the problem allows for $1/M'' \neq 0$
which implies an elliptic shape of the hole pockets as illustrated in figure 4.
\begin{figure}[t]
\begin{center}
\vspace{-0.4cm}
\epsfig{file=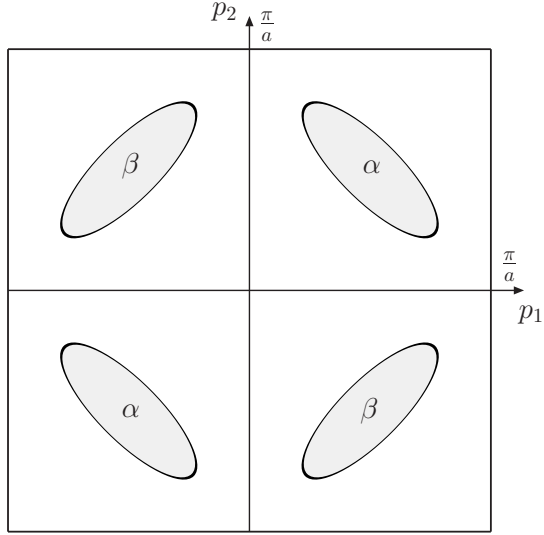,width=8cm}
\end{center}
\caption{\it Elliptically shaped hole pockets centered at 
$(\pm \frac{\pi}{2a},\pm \frac{\pi}{2a})$. Two pockets centered at $k^f$
and ${k^f}'$ combine to form the pockets for the flavors $f = \alpha,\beta$.}
\end{figure}
This is indeed observed both in ARPES experiments and in numerical simulations 
of $t$-$J$-type models (see figure 1). It should be noted that stability of the
minima at the center of the hole pockets requires $|M''| > M'$.

The leading terms with four fermion fields are given by
\begin{align}
{\cal L}_4 = \sum_{s = +,-} \Big\{ &
\frac{G_1}{2} (\psi^{\alpha\dagger}_s \psi^\alpha_s 
\psi^{\alpha\dagger}_{-s} \psi^\alpha_{-s} + 
\psi^{\beta\dagger}_s \psi^\beta_s 
\psi^{\beta\dagger}_{-s} \psi^\beta_{-s}) \nonumber \\[-1ex]
&+G_2 \psi^{\alpha\dagger}_s \psi^\alpha_s \psi^{\beta\dagger}_s \psi^\beta_s +
G_3 \psi^{\alpha\dagger}_s \psi^\alpha_s 
\psi^{\beta\dagger}_{-s} \psi^\beta_{-s} \nonumber \\[.7ex]
&+G_4 \Big[ \psi^{\alpha\dagger}_s \psi^\alpha_s
\sum_{s' = +,-} \big( \psi^{\beta\dagger}_{s'} v^{s'}_1 \psi^\beta_{-s'} - 
\psi^{\beta\dagger}_{s'} v^{s'}_2 \psi^\beta_{-s'} \big) \nonumber \\
&\hspace{2.7em}+\psi^{\beta\dagger}_s \psi^\beta_s
\sum_{s' = +,-} \big( \psi^{\alpha\dagger}_{s'} v^{s'}_1 \psi^\alpha_{-s'} + 
\psi^{\alpha\dagger}_{s'} v^{s'}_2 \psi^\alpha_{-s'} \big) \Big] \Big\},
\end{align}
with the real-valued 4-fermion contact interactions $G_1$, $G_2$, $G_3$, and 
$G_4$. We have limited ourselves to terms containing at most one spatial
derivative. The next order contains a large number of terms with one temporal 
or two spatial derivatives. We have constructed these terms using the algebraic
program FORM \cite{Ver00}, but we do not list them here because they are not 
very illuminating. Also, due to the large number of low-energy parameters they 
contain, they are unlikely to be used in any practical investigation.

For completeness, we also list the contributions to the Lagrangian with six and
eight fermion fields
\begin{align}
{\cal L}_6=&\,
H \big( \psi^{\alpha\dagger}_+ \psi^\alpha_+ \psi^{\alpha\dagger}_- 
\psi^\alpha_- \psi^{\beta\dagger}_+ \psi^\beta_+ +
\psi^{\alpha\dagger}_+ \psi^\alpha_+ \psi^{\alpha\dagger}_- \psi^\alpha_-
\psi^{\beta\dagger}_- \psi^\beta_- \nonumber \\
&\hspace{1.3em}+\psi^{\alpha\dagger}_+ \psi^\alpha_+ \psi^{\beta\dagger}_+
\psi^\beta_+ \psi^{\beta\dagger}_- \psi^\beta_- +
\psi^{\alpha\dagger}_- \psi^\alpha_- \psi^{\beta\dagger}_+ \psi^\beta_+
\psi^{\beta\dagger}_- \psi^\beta_- \big), \nonumber \\[0.5ex]
{\cal L}_8=&\,
I \, \psi^{\alpha\dagger}_+ \psi^\alpha_+ \psi^{\alpha\dagger}_- \psi^\alpha_- 
\psi^{\beta\dagger}_+ \psi^\beta_+ \psi^{\beta\dagger}_- \psi^\beta_-.
\end{align}
Here we have limited ourselves to terms without derivatives. Terms with more
fermion fields are then excluded by the Pauli principle. Again, it is 
straightforward to systematically construct the higher-order terms, but there 
is presently no need for them.

\subsection{Accidental Emergent Symmetries}

Interestingly, the terms in the Lagrangian constructed above have an accidental
global $U(1)_F$ flavor symmetry that acts as
\begin{equation}
U(1)_F: \quad ^F\psi^f_\pm(x) = \exp(\sigma_f i \eta) \psi^f_\pm(x).
\end{equation}
The flavor symmetry is explicitly broken by higher-order terms in the 
derivative expansion and thus emerges only at low energies.

In addition, for $c \rightarrow \infty$ there is also an accidental Galilean 
boost symmetry $G$, which acts on the fields as
\begin{alignat}{2}
\label{boost}
G:&\quad &^GP(x) &= P(Gx), \qquad Gx = (\vec x - \vec v \ t,t), \nonumber \\
  &\quad &^G \psi^f_\pm(x) &= 
  \exp\left(i \vec p^f \cdot \vec x - \omega^f t\right) \psi^f_\pm(Gx),
\nonumber \\
  &\quad &^G \psi^{f\dagger}_\pm(x) &= 
  \psi^{f\dagger}_\pm(Gx) 
\exp\left(- i \vec p^f \cdot \vec x + \omega^f t\right),
\end{alignat}
with $\vec p^f = (p^f_1, p^f_2)$ and $\omega^f$ given by
\begin{align}
p^f_1 &= \frac{M'}{1 - (M'/M'')^2}
\left[ v_1 - \sigma_f \frac{M'}{M''} v_2 \right], \quad
p^f_2 = \frac{M'}{1 - (M'/M'')^2}
\left[ v_2 - \sigma_f \frac{M'}{M''} v_1 \right], \nonumber \\
\omega^f &= \frac{{p^f_i}^2}{2 M'} + \sigma_f \frac{p^f_1 p^f_2}{M''} =
\frac{M'}{1 - (M'/M'')^2}\left[\frac{1}{2}(v_1^2 + v_2^2) 
- \sigma_f \frac{M'}{M''} v_1 v_2\right].
\end{align}
Note that the relation between $\vec p^f$ and the velocity of the Galilean 
boost $\vec v$ results from the hole dispersion relation of 
eq.(\ref{dispersion}) using $v_i = d E^f/d p_i^f$. Also the Galilean boost
symmetry is explicitly broken at higher orders of the derivative expansion.

The fundamental physics underlying the actual cuprates is Galilean- or, in 
fact, even Poincar\'e-invariant. Poincar\'e symmetry is then spontaneously 
broken by the formation of a crystal lattice with phonons as the corresponding 
Nambu-Goldstone bosons. In the Hubbard or $t$-$J$ models the lattice is imposed
by hand, and Galilean symmetry is thus broken explicitly instead of 
spontaneously. In particular, there are no phonons in these models. Remarkably,
an accidental Galilean boost invariance still emerges dynamically at low 
energies. This has important physical consequences. In particular, without loss
of generality, the hole pairs to be investigated later, can be studied in their
rest frame. This is unusual for particles propagating on a lattice, because the
lattice represents a preferred rest frame (a condensed matter ``ether''). The 
accidental Galilean boost invariance may even break spontaneously, which is the
case in phases with spiral configurations of the staggered magnetization.

\section{Magnon-mediated Binding between Holes}

Our treatment of the forces between two holes is analogous to the effective 
theory for light nuclei \cite{Wei90,Kap98,Epe98,Bed98} in which one-pion 
exchange dominates the long-range forces. In this section we calculate the
one-magnon exchange potentials between holes and we solve the corresponding
two-hole Schr\"odinger equations. The one-magnon exchange potentials as well as
the solution of the Schr\"odinger equation for a hole pair of flavors
$\alpha$ and $\beta$ were already discussed in \cite{Bru05}. Here we present a 
more detailed derivation of these results and we extend the discussion to hole
pairs of the same flavor.

\subsection{One-Magnon Exchange Potentials between Holes}

We now calculate the one-magnon exchange potentials between holes of flavors 
$\alpha$ or $\beta$. For this purpose, we expand in the magnon fluctuations 
$m_1(x)$, $m_2(x)$ around the ordered staggered magnetization, i.e.
\begin{align}
\vec e(x) = \Big( \frac{m_1(x)}{\sqrt{\rho_s}}&,\,
\frac{m_2(x)}{\sqrt{\rho_s}},1 \Big) + {\cal O}(m^2) \nonumber \\
\Rightarrow \quad
v_\mu^\pm(x) &= \frac{1}{2 \sqrt{\rho_s}} \p_\mu
\big[ m_2(x) \pm i m_1(x) \big] + {\cal O}(m^3), \nonumber \\
v_\mu^3(x) &= \frac{1}{4 \rho_s}\big[m_1(x) \p_\mu m_2(x) -
m_2(x) \p_\mu m_1(x)\big] + {\cal O}(m^4).
\end{align}
Since vertices with $v_\mu^3(x)$ (contained in $D_\mu$) involve at least two 
magnons, one-magnon exchange results from vertices with $v_\mu^\pm(x)$ only. 
As a consequence, two holes can exchange a single magnon only if they have
antiparallel spins ($+$ and $-$), which are both flipped in the magnon-exchange
process. We denote the momenta of the incoming and outgoing holes by 
$\vec p_\pm$ and $\vec p_\pm \ \!\!\!\! ' \ $, respectively. The momentum 
carried
by the exchanged magnon is denoted by $\vec q$. We also introduce the total 
momentum $\vec P$ as well as the incoming and outgoing relative momenta 
$\vec p$ and $\vec p \ '$
\begin{gather}
\vec P = \vec p_+ + \vec p_- = \vec p_+ \ \!\!\!\! ' + \vec p_- \ \!\!\!\! ',
\nonumber \\
\vec p = \frac{1}{2}(\vec p_+ - \vec p_-), \qquad
\vec p \ ' = \frac{1}{2}(\vec p_+ \ \!\!\!\! ' - \vec p_- \ \!\!\!\! ').
\end{gather}
Momentum conservation then implies
\begin{equation}
\vec q = \vec p + \vec p \ '.
\end{equation}
It is straightforward to evaluate the Feynman diagram describing one-magnon 
exchange shown in figure 5. 
\begin{figure}[tb]
\begin{center}
\vspace{-0.4cm}
\epsfig{file=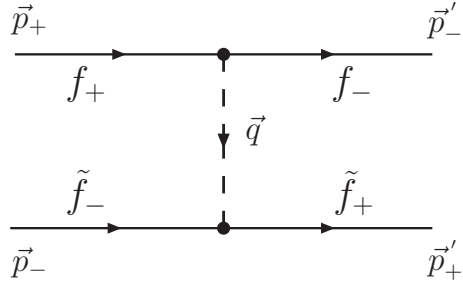,width=7cm}
\end{center}
\caption{\it Feynman diagram for one-magnon exchange between two holes with 
antiparallel spins undergoing a spin-flip.}
\end{figure}
In momentum space the resulting potentials for the various combinations of
flavors take the form
\begin{equation}
\langle \vec p_+ \ \!\!\!\! ' \vec p_- \ \!\!\!\! '|V^{f\tilde f}
|\vec p_+ \vec p_-\rangle = V^{f\tilde f}(\vec q) \
\delta(\vec p_+ + \vec p_- - \vec p_+ \ \!\!\!\! ' - \vec p_- \ \!\!\!\! '),
\end{equation}
with
\begin{gather}
V^{\alpha\alpha}(\vec q) = - \gamma \pi \frac{(q_1 + q_2)^2}{q^2},
\qquad
V^{\beta\beta}(\vec q) = - \gamma \pi \frac{(q_1 - q_2)^2}{q^2},
\nonumber \\
V^{\alpha\beta}(\vec q) = V^{\beta\alpha}(\vec q) = 
- \gamma \pi \frac{q_1^2 - q_2^2}{q^2},
\end{gather}
where $\gamma = \Lambda^2/(2 \pi \rho_s)$. In coordinate space the 
corresponding potentials are given by
\begin{equation}
\langle \vec r_+ \ \!\!\!\! ' \vec r_- \ \!\!\!\! '|V^{f\tilde f}
|\vec r_+ \vec r_-\rangle = V^{f\tilde f}(\vec r) \
\delta(\vec r_+ - \vec r_- \ \!\!\!\! ') \
\delta(\vec r_- - \vec r_+ \ \!\!\!\! '),
\end{equation}
with
\begin{gather}
V^{\alpha\alpha}(\vec r) = \gamma \frac{\sin(2 \varphi)}{r^2}, \qquad
V^{\beta\beta}(\vec r) = - \gamma \frac{\sin(2 \varphi)}{r^2},
\nonumber \\
V^{\alpha\beta}(\vec r) = V^{\beta\alpha}(\vec r) = 
\gamma \frac{\cos(2 \varphi)}{r^2}.
\end{gather}
Here $\vec r = \vec r_+ - \vec r_-$ is the distance vector between the two 
holes and $\varphi$ is the angle between $\vec r$ and the $x$-axis. It should 
be noted that the one-magnon exchange potentials are instantaneous although 
magnons travel with the finite speed $c$. Retardation effects occur only at 
higher orders. The one-magnon exchange potentials also contain short-distance 
$\delta$-function contributions which we have not listed above. These 
contributions add to the 4-fermion contact interactions. Since we will model 
the short-distance repulsion by a hard core radius, the $\delta$-function 
contributions are not needed in the following.

\subsection{Schr\"odinger Equation for two Holes of different Flavor}

Let us investigate the Schr\"odinger equation for the relative motion of two 
holes with flavors $\alpha$ and $\beta$. Thanks to the accidental Galilean 
boost invariance, without loss of generality we can consider the hole pair in 
its rest frame. The total kinetic energy of the two holes is then given by
\begin{equation}
T = \sum_{f = \alpha,\beta} \left(\frac{p_i^2}{2 M'} + 
\sigma_f \frac{p_1 p_2}{M''}\right) = \frac{p_i^2}{M'}. 
\end{equation}
In particular, the parameter $1/M''$ that measures the deviation from a 
circular shape of the hole pockets drops out of the problem. The resulting
Schr\"odinger equation then takes the form
\begin{equation}
\left(\begin{array}{cc} - \frac{1}{M'} \Delta & V^{\alpha\beta}(\vec r) 
\\[0.2ex]
V^{\alpha\beta}(\vec r) &  - \frac{1}{M'} \Delta \end{array} \right)
\left(\begin{array}{c} \Psi_1(\vec r) \\ 
\Psi_2(\vec r) \end{array}\right) = E 
\left(\begin{array}{c} \Psi_1(\vec r) \\ 
\Psi_2(\vec r) \end{array}\right).
\end{equation}
The components $\Psi_1(\vec r)$ and $\Psi_2(\vec r)$ are probability
amplitudes for the spin-flavor combinations $\alpha_+\beta_-$ and 
$\alpha_-\beta_+$, respectively. The potential $V^{\alpha\beta}(\vec r)$ 
couples the two channels because magnon exchange is accompanied by a spin-flip.
The above Schr\"odinger equation does not yet account for the short-distance 
forces arising from 4-fermion contact interactions. Their effect will be 
incorporated later by a boundary condition on the wave function near the 
origin. Making the ansatz 
\begin{equation}
\Psi_1(\vec r) \pm \Psi_2(\vec r) = R(r) \chi_\pm(\varphi),
\end{equation}
for the angular part of the wave function one obtains
\begin{equation}
- \frac{d^2\chi_\pm(\varphi)}{d\varphi^2} \pm 
M' \gamma \cos(2 \varphi) \chi_\pm(\varphi) = - \lambda \chi_\pm(\varphi).
\end{equation}
The solutions of this Mathieu equation with the lowest eigenvalue $- \lambda$ 
is
\begin{equation}
\chi_\pm(\varphi) = \frac{1}{\sqrt{\pi}}\,
\mbox{ce}_0 ( \varphi, \, \pm \frac{1}{2} M' \gamma ), \qquad 
\lambda = \frac{1}{8} (M' \gamma)^2 + {\cal O}(\gamma^4). 
\end{equation}
The periodic Mathieu function $\mbox{ce}_0(\varphi,\frac{1}{2} M' \gamma)$ 
\cite{Abr72} is illustrated in figure 6.
\begin{figure}[tb]
\begin{center}
\vspace{-0.3cm}
\epsfig{file=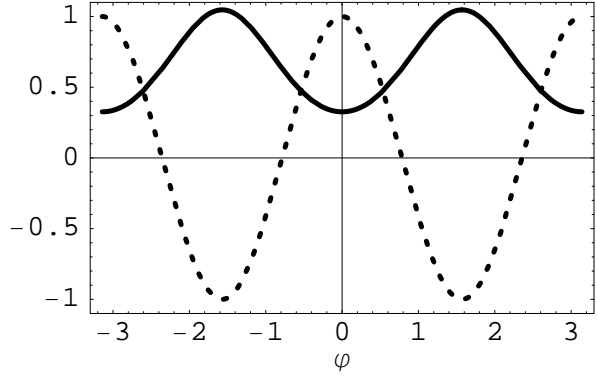,width=8cm}
\end{center}
\caption{\it Angular wave function $\mbox{ce}_0(\varphi,\frac{1}{2} M' \gamma)$
(solid curve) and angle-dependence $\cos(2 \varphi)$ of the potential (dotted 
curve) for a pair of holes with flavors $\alpha$ and $\beta$ 
($M' \gamma = 2.5$).}
\end{figure}

The radial Schr\"odinger equation takes the form
\begin{equation}
\label{radial}
- \left[\frac{d^2R(r)}{dr^2} + 
\frac{1}{r} \frac{dR(r)}{dr}\right] - \frac{\lambda}{r^2} R(r) = M' E R(r).
\end{equation}
As it stands, this equation is ill-defined because an attractive 
$\frac{1}{r^2}$ potential is too singular at the origin. However, we must still
incorporate the contact interaction proportional to the 4-fermion couplings
$G_3$ and $G_4$. A consistent description of the short-distance physics 
requires ultraviolet regularization and subsequent renormalization of the 
Schr\"odinger equation as discussed in \cite{Lep97}. In order to maintain
the transparency of a complete analytic calculation, here we model the 
short-distance repulsion between two holes by a hard core of radius $r_0$, 
i.e.\ we require $R(r_0) = 0$. The radial Schr\"odinger equation for the bound 
states is solved by a Bessel function
\begin{equation}
R(r) = A K_\nu \big( \sqrt{M' |E_n|} r \big), \qquad \nu = i \sqrt{\lambda}.
\end{equation}
The energy (determined from $K_\nu \big( \sqrt{M' |E_n|} r_0 \big) = 0$)
is given by
\begin{equation}
E_n \sim - (M' r_0^2)^{-1} \exp(- 2 \pi n/\sqrt{\lambda})
\end{equation}
for large $n$. As expected, the energy of the bound state depends on the values
of the low-energy constants. Although the binding energy is exponentially small
in $n$, for very small $r_0$ the ground state would have a small size and would
be strongly bound. In that case, the result of the effective theory should not 
be trusted quantitatively, because short-distance details and not the universal
magnon-dominated long-distance physics determine the structure of the bound 
state. Still, even in that case, an extended effective theory can be 
constructed which contains the tightly bound hole pairs as additional explicit 
low-energy degrees of freedom. For larger values of $r_0$, as long as the 
binding energy is small compared to the relevant high-energy scales such as 
$\rho_s$, the results of the effective theory in its present form are reliable,
and receive only small calculable corrections from higher-order effects such as
two-magnon exchange.

It should be noted that the wave functions with angular part $\chi_+(\varphi)$ 
and $\chi_-(\varphi)$ have the same energy. A general linear combination of the
two states takes the form
\begin{equation}
\Psi(\vec r) = R(r) \left(\begin{array}{c}
a \chi_+(\varphi) + b \chi_-(\varphi) \\
a \chi_+(\varphi) - b \chi_-(\varphi) \end{array} \right).
\end{equation}
Applying the 90 degrees rotation $O$ and using the transformation rules of 
eq.(\ref{symcomp}) one obtains
\begin{equation}
^O\Psi(\vec r) = R(r) \left(\begin{array}{c}
a \chi_+(\varphi + \frac{\pi}{2}) - b \chi_-(\varphi + \frac{\pi}{2}) \\
- a \chi_+(\varphi + \frac{\pi}{2}) - b \chi_-(\varphi + \frac{\pi}{2}) 
\end{array} \right) =
R(r) \left(\begin{array}{c}
a \chi_-(\varphi) - b \chi_+(\varphi) \\
- a \chi_-(\varphi) - b \chi_+(\varphi) \end{array} \right).
\end{equation}
Demanding that $\Psi(\vec r)$ is an eigenstate of the rotation $O$, i.e.\
$^O\Psi(\vec r) = o \Psi(\vec r)$, thus implies
\begin{equation}
o \ a = - b, \quad o \ b = a \ \Rightarrow \ o = \pm i,
\end{equation}
with the corresponding eigenfunctions given by
\begin{equation}
\Psi_\pm(\vec r) = R(r) \left(\begin{array}{c}
\chi_+(\varphi) \mp i \chi_-(\varphi) \\
\chi_+(\varphi) \pm i \chi_-(\varphi) \end{array} \right).
\end{equation}
This leads to the probability distribution illustrated in figure 7, which 
resembles $d_{x^2-y^2}$ symmetry. However, unlike for a true $d$-wave, the wave
function is suppressed, but not equal to zero, along the lattice diagonals.
This is different for the first angular-excited state, whose wave function 
indeed has a node along the diagonals.
\begin{figure}[tb]
\begin{center}
\vspace{-0.3cm}
\epsfig{file=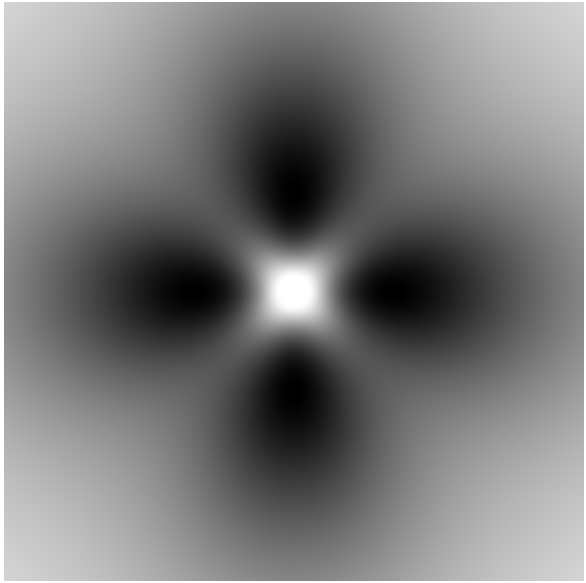,width=8cm}
\end{center}
\caption{\it Probability distribution for the ground state of two holes with 
flavors $\alpha$ and $\beta$.}
\end{figure}
Since the problem only has a 90 degrees and not a continuous rotation symmetry,
the continuum classification scheme of angular momentum eigenstates does not 
apply here. In fact, the 2-fold degenerate ground state belongs to the 
2-dimensional irreducible representation of the group of discrete rotations
and reflections. The corresponding eigenvalues of the 90 degrees rotation $O$
are $o = \pm i$.

It is also interesting to investigate the transformation properties under the
reflection symmetry $R$ and the unbroken shift symmetries $D_i'$. Under the
reflection $R$ one obtains
\begin{equation}
^R\Psi_\pm(\vec r) = R(r) \left(\begin{array}{c}
\chi_+(- \varphi) \pm i \chi_-(- \varphi) \\
\chi_+(- \varphi) \mp i \chi_-(- \varphi) \end{array} \right) =
R(r) \left(\begin{array}{c}
\chi_+(\varphi) \pm i \chi_-(\varphi) \\
\chi_-(\varphi) \mp i \chi_+(\varphi) \end{array} \right) = \Psi_\mp(\vec r). 
\end{equation}
Similarly, under the displacement symmetries one obtains
\begin{eqnarray}
&&^{D_1'}\Psi_\pm(\vec r) = R(r) \left(\begin{array}{c}
\chi_+(\varphi) \pm i \chi_-(\varphi) \\
\chi_+(\varphi) \mp i \chi_-(\varphi) \end{array} \right) = \Psi_\mp(\vec r),
\nonumber \\
&&^{D_2'}\Psi_\pm(\vec r) = - R(r) \left(\begin{array}{c}
\chi_+(\varphi) \pm i \chi_-(\varphi) \\
\chi_+(\varphi) \mp i \chi_-(\varphi) \end{array} \right) = - \Psi_\mp(\vec r).
\end{eqnarray}

\subsection{Schr\"odinger Equation for two Holes of the same Flavor}

Let us now consider two holes of the same flavor. In particular, we focus on an
$\alpha \alpha$ pair. Hole pairs of type $\beta \beta$ behave in exactly the 
same way. For simplicity, we first consider the (somewhat unrealistic) case of 
circular hole pockets. Then we discuss the realistic (but slightly more 
complicated) case of elliptically shaped pockets. 

\subsubsection{Circular Hole Pockets}

\begin{figure}[tb]
\begin{center}
\vspace{-0.3cm}
\epsfig{file=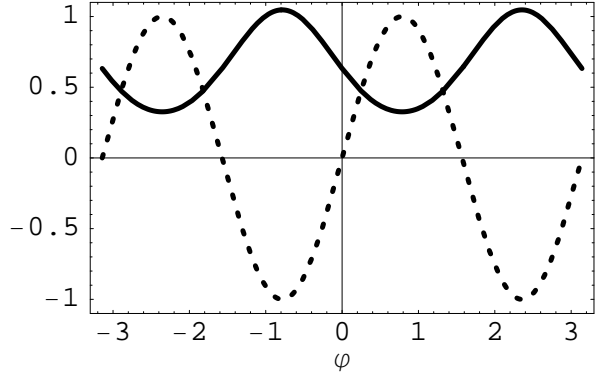,width=8cm}
\end{center}
\caption{\it Angular wave function 
$\mbox{ce}_0(\varphi - \frac{\pi}{4},\frac{1}{2} M' \gamma)$
(solid curve) and angle-dependence $\sin(2 \varphi)$ of the potential (dotted 
curve) for two holes of flavor $\alpha$ residing in a circular hole pocket
($M' \gamma = 2.5$).}
\end{figure}
Let us consider two holes of flavor $\alpha$ with opposite
spins $+$ and $-$. In the rest frame the wave function depends on the relative
distance vector $\vec r$ which points from the spin $+$ hole to the spin $-$ 
hole. It is important to note that magnon exchange is accompanied by a 
spin-flip. Hence, the vector $\vec r$ changes its direction in the magnon 
exchange process. For circular hole pockets, i.e.\ for $1/M'' = 0$, the total 
kinetic energy is again given by $T = p_i^2/M'$ and the resulting 
Schr\"odinger equation takes the form
\begin{equation}
- \frac{1}{M'} \Delta \Psi(\vec r) + V^{\alpha\alpha}(\vec r) \Psi(- \vec r) =
E \Psi(\vec r).
\end{equation}
As before, we make a separation ansatz
\begin{equation}
\Psi(\vec r) = R(r) \chi(\varphi).
\end{equation}
We concentrate on the ground state which is even with respect to the reflection
of $\vec r$ to $- \vec r$, i.e.\
\begin{equation}
\chi(\varphi + \pi) = \chi(\varphi).
\end{equation}
The angular part of the Schr\"odinger equation then reads
\begin{equation}
- \frac{d^2\chi(\varphi)}{d\varphi^2} + 
M' \gamma \sin(2 \varphi) \chi(\varphi) = - \lambda \chi(\varphi).
\end{equation}
Again, this is a Mathieu equation. The ground state with eigenvalue $- \lambda$
takes the form
\begin{equation}
\chi(\varphi) = \frac{1}{\sqrt{\pi}}
\mbox{ce}_0 \big( \varphi - \frac{\pi}{4},\,\frac{1}{2} M' \gamma \big), \qquad
\lambda = \frac{1}{8} (M' \gamma)^2 + {\cal O}(\gamma^4).
\end{equation}
The angular wave function for the ground state together with the angular 
dependence of the one-magnon exchange potential are shown in figure 8.

As before, the radial Schr\"odinger equation takes the form of 
eq.(\ref{radial}). Again, we model the short-distance repulsion between two 
holes by a hard core of radius $r_0'$, i.e.\ we require $R(r_0') = 0$. It 
should be noted that $r_0'$ does not necessarily take the same value as $r_0$ 
in the $\alpha \beta$ case. This is not only because there is an additional 
$\delta$-function contribution to the one-magnon exchange potential, but also 
because the 4-fermion coupling $G_1$ in the $\alpha \alpha$ case is in general 
different from the coupling $G_3$ in the $\alpha \beta$ case. The energy is 
then given by
\begin{equation}
E_n \sim - (M' {r_0'}^2)^{-1} \exp(- 2 \pi n/\sqrt{\lambda})
\end{equation}
for large $n$. Again, there are two degenerate ground states --- one for an
$\alpha \alpha$ and one for a $\beta \beta$ pair. They are eigenstates of 
flavor related to each other by a 90 degrees rotation. Since the $U(1)_F$ 
symmetry is accidental at low energies while the 90 degrees rotation symmetry 
is exact, it is again natural to combine the two degenerate states to 
eigenstates of the rotation symmetry $O$. The resulting probability 
distribution which resembles $d_{xy}$ symmetry is illustrated in figure 9.
\begin{figure}[tb]
\begin{center}
\vspace{-0.3cm}
\epsfig{file=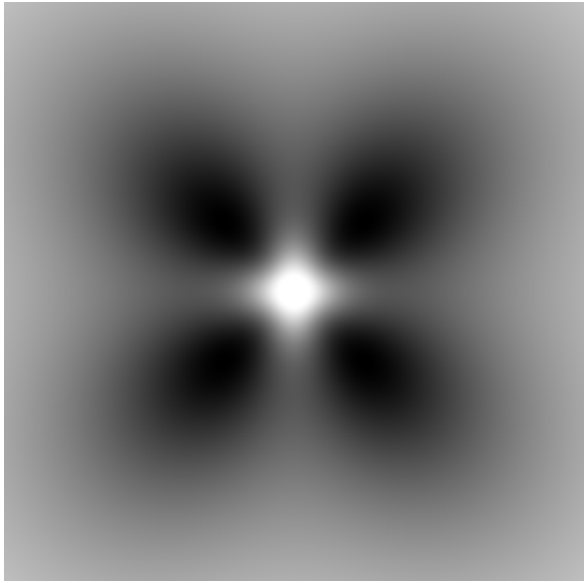,width=8cm}
\end{center}
\caption{\it Probability distribution for the ground state of two holes with 
flavors $\alpha \alpha$ or $\beta \beta$, combined to an eigenstate of the 90
degrees rotation symmetry $O$, for the case of circular hole pockets.}
\end{figure}
As for $\alpha \beta$ pairs, the symmetry is not truly $d$-wave, but just
given by the 2-dimensional irreducible representation of the group of discrete
rotations and reflections. Again, the corresponding eigenvalues of the 90 
degrees rotation $O$ are $o = \pm i$.

\subsubsection{Elliptic Hole Pockets}

Let us now move on to the realistic case of elliptically shaped hole pockets.
Then the total kinetic energy of two holes of flavor $\alpha$ in their rest 
frame is given by
\begin{equation}
T = \frac{p_i^2}{M'} + \frac{2 p_1 p_2}{M''}.
\end{equation}
This suggests to rotate the coordinate system by 45 degrees such that the major
axes of the ellipse are aligned with the rotated coordinate axes, i.e.
\begin{equation}
p_1' = \frac{1}{\sqrt{2}}(p_1 + p_2), \qquad
p_2' = \frac{1}{\sqrt{2}}(p_1 - p_2).
\end{equation}
In the rotated reference frame, the kinetic energy takes the form
\begin{equation}
T = \frac{{p_1'} ^2}{M_1} + \frac{{p_2'} ^2}{M_2},
\end{equation}
with
\begin{equation}
\frac{1}{M_1} = \frac{1}{M'} + \frac{1}{M''}, \qquad
\frac{1}{M_2} = \frac{1}{M'} - \frac{1}{M''}.
\end{equation} 
It is convenient to rescale the rotated axes such that the hole pocket again
assumes a circular shape. This is achieved by defining
\begin{equation}
\tild p_1 = \sqrt{\frac{M'}{M_1}} p_1', \qquad
\tild p_2 = \sqrt{\frac{M'}{M_2}} p_2',
\end{equation}
which indeed implies
\begin{equation}
T = \frac{\tild p_1^2}{M'} + \frac{\tild p_2^2}{M'} = 
\frac{\tild p_i^2}{M'},
\end{equation}
just as for the circular hole pocket. Of course, the rotation and rescaling 
must also be applied to the coordinates, i.e.
\begin{equation}
\tild x_1 = \sqrt{\frac{M_1}{2M'}}(x_1 + x_2) = \tild r \cos \tild \varphi, 
\qquad
\tild x_2 = \sqrt{\frac{M_2}{2M'}}(x_1 - x_2) = \tild r \sin \tild \varphi.
\end{equation}
The rotated and rescaled one-magnon exchange potential then takes the form
\begin{eqnarray}
V^{\alpha\alpha}(\tild{\vec r})&=&\gamma \frac{\sin(2 \varphi)}{r^2} =
\gamma \frac{2 x_1 x_2}{(x_1^2 + x_2^2)^2} =
\gamma \frac{\tild x_1^2 \, M'/M_1 - \tild x_2^2 \, M'/M_2}
{(\tild x_1^2 \, M'/M_1 + \tild x_2^2 \, M'/M_2)^2} \nonumber \\
&=&\gamma \frac{\cos(2 \tild \varphi) + M'/M''}
{{\tild r}^2 (1 + \cos(2 \tild \varphi) \, M'/M'')^2},
\end{eqnarray}
and the corresponding Schr\"odinger equation reads
\begin{equation}
- \frac{1}{M'} \Delta \Psi(\tild{\vec r}) + 
V^{\alpha\alpha}(\tild{\vec r}) \Psi(- \tild{\vec r}) = 
E \Psi(\tild{\vec r}).
\end{equation}
Once again, we make the separation ansatz
\begin{equation}
\Psi(\tild{\vec r}) = R(\tild r) \chi(\tild \varphi),
\end{equation}
such that the angular part of the Schr\"odinger equation now takes the form
\begin{equation}
- \frac{d^2\chi(\tild \varphi)}{d\tild \varphi^2} + 
M' \gamma \frac{\cos(2 \tild \varphi) + M'/M''}
{(1 + \cos(2 \tild \varphi) \, M'/M'')^2} \chi(\tild \varphi) = 
- \lambda \chi(\tild \varphi).
\end{equation}
This is a differential equation in the class of Hill equations \cite{Mag66}
which we have solved numerically. Figure 10 shows the angular wave function for
the ground state together with the angular dependence of the rotated and 
rescaled one-magnon exchange potential.
\begin{figure}[tb]
\begin{center}
\vspace{-0.3cm}
\epsfig{file=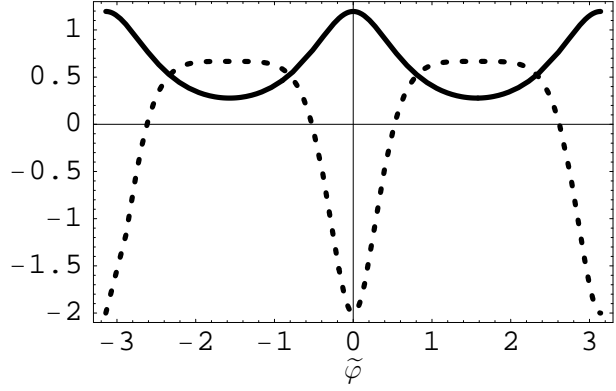,width=8cm}
\end{center}
\caption{\it Angular wave function (solid curve) and angle-dependence of the
rotated and rescaled one-magnon exchange potential (dotted curve) for two holes
of flavor $\alpha$ in an elliptic hole pocket (for $M'/M'' = 0.5$ and 
$M' \gamma = 2.5$).}
\end{figure}
The radial Schr\"odinger equation takes exactly the same form as for circular
hole pockets and will therefore not be discussed again. There are two 
degenerate states corresponding to $\alpha \alpha$ and $\beta \beta$ pairs, 
which are related to each other by a 90 degrees rotation. Combining the two
degenerate states to an eigenstate of the rotation $O$, one obtains the
probability distribution of figure 11 which again resembles $d_{xy}$ symmetry.
\begin{figure}[tb]
\begin{center}
\vspace{-0.3cm}
\epsfig{file=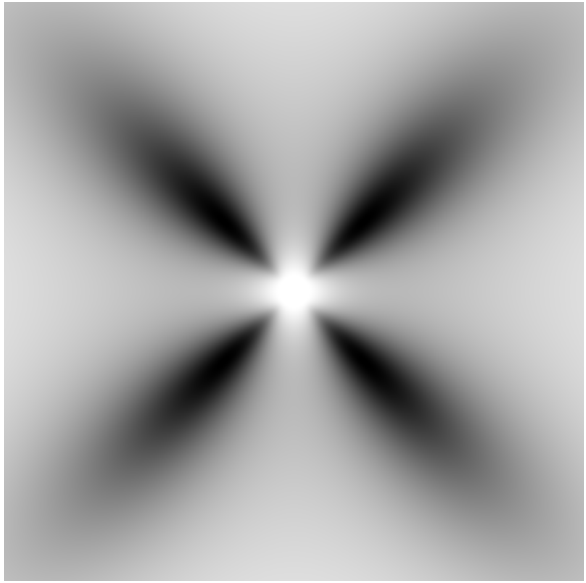,width=8cm}
\end{center}
\caption{\it Probability distribution for the ground state of two holes with 
flavors $\alpha \alpha$ or $\beta \beta$ in elliptic hole pockets, linearly 
combined to an eigenstate of the 90 degrees rotation $O$.}
\end{figure}
Qualitatively similar results for hole pairs from the same hole pocket were 
obtained in \cite{Kuc93} directly from the $t$-$J$ model, however, using some 
uncontrolled approximations. The effective field theory treatment has the 
advantage of being systematic, i.e.\ it can be improved order by order in the 
derivative expansion. Besides this, the effective field theory approach is 
particularly transparent and conceptually simple.

\subsection{Low-Energy Dynamics of Hole Pairs}

As we have seen, magnons mediate attractive forces between holes of opposite
spin which leads to bound states with $d$-wave characteristics. Once pairs of 
holes have formed, it is natural to ask how they behave at low temperatures. It
should be stressed that, until now, we have considered isolated hole pairs in 
an otherwise undoped antiferromagnet. What should one expect when the system is
doped with a non-zero density of holes? First of all, even for infinitesimal 
doping, the antiferromagnet may become unstable against the formation of
inhomogeneities, such as spirals in the staggered magnetization \cite{Shr88}. 
This phenomenon can be studied within the effective theory and is presently 
under investigation \cite{Bru06}. It turns out that, for sufficiently large 
spin stiffness $\rho_s$, the homogeneous antiferromagnet is stable, while for 
intermediate values of $\rho_s$ it becomes unstable against the formation of a
spiral phase. For even smaller values of $\rho_s$, the spiral itself becomes
unstable against the formation of further, yet unidentified, inhomogeneities.
In the following, we will assume a sufficiently large value of $\rho_s$ such
that the homogeneous antiferromagnet is stable.

If the short-range repulsion is particularly strong in the $\alpha\alpha$ and 
$\beta\beta$ channels, hole pairs of type $\alpha\beta$ with 
$d_{x^2 - y^2}$-like symmetry are most strongly bound. On the other hand, if 
the short-range repulsion is stronger in the $\alpha\beta$ channel, 
$\alpha\alpha$ or $\beta\beta$ pairs form and the symmetry resembles $d_{xy}$. 
At very low densities and temperatures, the hole pairs form a dilute system of 
bosons. In this case, the wave functions of different pairs do not overlap 
substantially, and it is natural to assume that they may undergo Bose-Einstein 
condensation at sufficiently low temperatures. At larger densities, the wave 
functions of different pairs begin to overlap. In that case, a momentum space 
description of holes is more appropriate. Magnon exchange between holes near 
the Fermi surface is then likely to produce an instability against Cooper pair 
formation in the $d$-wave channel. This is expected to lead to BCS-type 
superconductivity, mediated by magnons instead of phonons. Using the effective 
theory, the critical temperature for magnon-mediated BCS-type superconductivity
will be calculated elsewhere, but is not expected to be very high. In 
particular, superconductivity coexisting with antiferromagnetism at low doping 
is not observed in the cuprates. This may well be due to impurities on which 
holes get localized, thus preventing superconductivity. Magnon-mediated 
superconductivity in lightly doped antiferromagnets can still be studied 
systematically using the effective field theory which does not contain 
impurities.

\section{Conclusions}

Based on a careful symmetry analysis of the Hubbard and $t$-$J$ models, we have
constructed a systematic low-energy effective field theory for magnons and 
holes in lightly doped antiferromagnets. The effective theory provides a 
powerful framework in which nontrivial aspects of the strongly coupled dynamics
of the cuprates, 
such as the magnon-mediated forces between holes, can be addressed using 
systematic methods of weak coupling perturbation theory. The effective theory 
relies on a number of basic assumptions. 
Besides fundamental principles of field theory, such as locality, unitarity, 
and symmetry, it is based on the assumption that the $SU(2)_s$ spin symmetry is
spontaneously broken down to $U(1)_s$. This assumption is very accurately 
verified both in experiments with the cuprates and by numerical calculations in
the Hubbard and $t$-$J$ model at half-filling. A second basic assumption is 
that fermionic quasi-particles --- the holes of the effective theory --- indeed
exist as stable excitations located in specific pockets in the Brillouin zone. 
The location of these pockets in momentum space has been obtained from ARPES 
measurements as well
as from numerical computations in Hubbard and $t$-$J$ models. It should be 
stressed that the applicability of the effective theory depends crucially on 
the question if the relevant low-energy degrees of freedom have indeed been 
identified correctly. Since it is impossible to rigorously derive the effective
theory from an underlying microscopic system, we can not yet be completely 
sure that our effective theory indeed describes their low-energy physics 
correctly. In order to verify the correctness and accuracy of the effective 
theory, besides the theoretical considerations presented here, it will be 
important to confront it with experimental or numerical data. For undoped 
antiferromagnets, such confrontation led to a spectacular quantitative 
success of the pure magnon effective theory \cite{Wie94,Bea96}. It is expected 
that the full effective theory including holes will be equally successful. An 
important next step will be the comparison with precise Monte Carlo data, 
e.g.\, for the $t$-$J$ model. This will allow us to fix the values of the 
low-energy parameters of the effective theory in terms of the microscopic 
parameters $t$ and $J$. 

The effective theory can be used to calculate a wide variety of physical 
processes. The most basic processes include, for example, magnon-magnon and 
magnon-hole scattering. In order to correct the leading-order tree-level 
diagrams by a systematic loop expansion, in analogy to chiral perturbation 
theory, a power-counting scheme must be constructed. This is straightforward in
the pure-magnon effective theory and should be generalized to the full 
effective theory including charge carriers along the lines of \cite{Bec99}.
We have performed a systematic effective field theory analysis of the 
magnon-mediated forces between holes in an antiferromagnet. One-magnon exchange
mediates forces exclusively between holes of opposite spin. The leading terms 
in the fermionic part of the effective action are Galilean boost invariant and 
the two-hole system can thus be studied in its rest frame. Remarkably, some of 
the resulting two-particle Schr\"odinger equations can be solved completely 
analytically. Hole-doped cuprates have hole pockets centered at lattice momenta
$k^\alpha = (\frac{\pi}{2a},\frac{\pi}{2a})$ and
$k^\beta = (\frac{\pi}{2a},- \frac{\pi}{2a})$. As a consequence, the holes 
carry a ``flavor''-index $f = \alpha,\beta$ which specifies the pocket to which
a hole belongs. At leading order,
flavor is a conserved quantum number, and one can thus 
distinguish hole pairs of the types $\alpha\alpha$ or $\beta\beta$ from those 
of type $\alpha\beta$. Magnon exchange occurs with the same strength for both 
types, and leads to bound hole pairs with $d$-wave symmetry. For pairs of type 
$\alpha\alpha$ or $\beta\beta$ the symmetry resembles $d_{xy}$, while for those
of type $\alpha\beta$ it is $d_{x^2 - y^2}$-like. Depending on the strength of 
the respective short-distance repulsion either the pairs of type $\alpha\alpha$
and $\beta\beta$ or those of type $\alpha\beta$ are more strongly bound. At low
densities and temperatures, the hole pairs may undergo Bose-Einstein
condensation. Once the wave functions of different pairs overlap substantially,
one may expect BCS-type magnon-mediated $d$-wave superconductivity coexisting 
with antiferromagnetism. Although coexistence of antiferromagnetism and 
superconductivity is not observed in the cuprates --- possibly due to the 
localization of holes on impurities --- the exchange of spin fluctuations is a 
promising potential mechanism for high-temperature superconductivity. In 
lightly doped systems without impurities, magnon-mediated superconductivity
can be studied systematically using the effective theory. Other applications of
the effective theory, which are currently under investigation, aim at a 
quantitative understanding of the destruction of antiferromagnetism upon doping
and of the generation of spiral phases \cite{Bru06}.

It is also natural to ask if the effective theory can possibly be applied to 
the high-temperature superconductors themselves. Since in the real materials,
which contain impurities, high-temperature superconductivity arises only after
antiferromagnetism has been destroyed, and since the effective theory relies on
the spontaneous breakdown of the $SU(2)_s$ symmetry, this seems doubtful. 
However, while the systematic treatment of the effective theory will break down
in the superconducting phase, the effective theory itself does not, as long as 
spin fluctuations (now with a finite correlation length) and holes in momentum 
space pockets $(\frac{\pi}{2a},\pm \frac{\pi}{2a})$ remain the relevant degrees
of freedom. After all, the effective theory of magnons and holes can also be 
considered beyond perturbation theory, for example, by regularizing it on a 
lattice and simulating it numerically. A similar procedure has been discussed 
for the effective theory of pions and nucleons \cite{Cha03}. Unfortunately, one
would expect that the sign problem will once more raise its ugly head, and may 
thus prevent efficient numerical simulations not only in the microscopic models
but also in the effective theory. It thus remains to be seen if nonperturbative
investigations of the effective theory can shed light on the phenomenon of 
high-temperature superconductivity itself. We prefer to first concentrate on 
lightly doped systems without impurities for which the low-energy effective 
field theory makes quantitative predictions. Once these idealized systems are 
better understood, further steps towards understanding the more complicated 
actual materials can be taken on a more solid theoretical basis.

\section*{Acknowledgements}

This work is supported by funds provided by the Schweizerischer Nationalfonds
as well as by the INFN.


\begin{thebibliography}{10}

\bibitem{Bed86}
J.\ C.\ Bednorz and K.\ A.\ M\"uller, Z.\ Phys.\ B64 (1986) 189.

\bibitem{Bri70}
W.\ F.\ Brinkman and T.\ M.\ Rice, Phys.\ Rev.\ B2 (1970) 1324.

\bibitem{Hir85}
J.\ E.\ Hirsch, Phys.\ Rev.\ Lett.\ 54 (1985) 1317.

\bibitem{And87}
P.\ W.\ Anderson, Science 235 (1987) 1196.

\bibitem{Gro87}
C.\ Gros, R.\ Joynt, and T.\ M.\ Rice, Phys.\ Rev.\ B36 (1987) 381.

\bibitem{Shr88}
B.\ I.\ Shraiman and E.\ D.\ Siggia, Phys.\ Rev.\ Lett.\ 60 (1988) 740;
Phys.\ Rev.\ Lett.\ 61 (1988) 467; Phys.\ Rev.\ Lett.\ 62 (1989) 1564;
Phys.\ Rev.\ B46 (1992) 8305.

\bibitem{Tru88}
S.\ A.\ Trugman, Phys.\ Rev.\ B37 (1988) 1597.

\bibitem{Sch88}
J.\ R.\ Schrieffer, X.\ G.\ Wen, and S.\ C.\ Zhang, Phys.\ Rev.\ Lett.\ 60
(1988) 944; Phys.\ Rev.\ B39 (1989) 11663.

\bibitem{Kan89}
C.\ L.\ Kane, P.\ A.\ Lee, and N.\ Read, Phys.\ Rev.\ B39 (1989) 6880.

\bibitem{Sac89}
S.\ Sachdev, Phys.\ Rev.\ B39 (1989) 12232.

\bibitem{Wen89}
X.~G.~Wen, Phys.\ Rev.\ B39 (1989) 7223.

\bibitem{Kra89}
A.\ Krasnitz, E.\ G.\ Klepfish, and A.\ Kovner, Phys.\ Rev.\ B39 (1989) 9147.

\bibitem{Sha90}
R.~Shankar, Phys.\ Rev.\ Lett.\ 63 (1989) 203; Nucl.\ Phys.\ B330 (1990) 433.

\bibitem{And90}
P.\ W.\ Anderson, Phys.\ Rev.\ Lett.\ 64 (1990) 1839.

\bibitem{Sin90}
A.\ Singh and Z.\ Tesanovic, Phys.\ Rev.\ B41 (1990) 614.

\bibitem{Tru90}
S.\ A.\ Trugman, Phys.\ Rev.\ B41 (1990) 892.

\bibitem{Els90}
V.\ Elser, D.\ A.\ Huse, B.\ I.\ Shraiman, and E.\ D.\ Siggia, Phys.\ Rev.\ B41
(1990) 6715.

\bibitem{Dag90}
E.\ Dagotto, R.\ Joynt, A.\ Moreo, S.\ Bacci, and E.\ Gagliano, 
Phys.\ Rev.\ B41 (1990) 9049.

\bibitem{Vig90}
G.\ Vignale and M.\ R.\ Hedayati, Phys.\ Rev.\ B42 (1990) 786.

\bibitem{Kop90}
P.\ Kopietz, Phys.\ Rev.\ B42 (1990) 1029.

\bibitem{Sch90}
H.\ J.\ Schulz, Phys.\ Rev.\ Lett.\ 65 (1990) 2462.

\bibitem{Wen91}
Z.\ Y.\ Weng, C.\ S.\ Ting, and T.\ K.\ Lee, Phys.\ Rev.\ B43 (1991) 3790.

\bibitem{Ver91}
J.\ A.\ Verges, E.\ Louis, P.\ S.\ Lomdahl, F.\ Guinea, and A.\ R.\ Bishop,
Phys.\ Rev.\ B43 (1991) 6099.

\bibitem{Mar91}
F.\ Marsiglio, A.\ E.\ Ruckenstein, S.\ Schmitt-Rink, and C.\ M.\ Varma,
Phys.\ Rev.\ B43 (1991) 10882.

\bibitem{Mon91}
P.\ Monthoux, A.\ V.\ Balatsky, and D.\ Pines, Phys.\ Rev.\ Lett.\ 67 (1991)
3448.

\bibitem{Liu92}
Z.\ Liu and E.\ Manousakis, Phys.\ Rev.\ B45 (1992) 2425.

\bibitem{Kue93}
C.\ K\"ubert and A.\ Muramatsu, Phys.\ Rev.\ B47 (1993) 787; cond-mat/9505105.

\bibitem{Dah93}
T.\ Dahm, J.\ Erdmenger, K.\ Scharnberg, and C.\ T.\ Rieck, Phys.\ Rev.\ B48
(1993) 3896.

\bibitem{Kuc93}
M.\ Y.\ Kuchiev and O.\ P.\ Sushkov, Physica C218 (1993) 197.

\bibitem{Fla93}
V.\ V.\ Flambaum, M.\ Y.\ Kuchiev, and O.\ P.\ Sushkov, Physica C227 (1994) 
467.

\bibitem{Sus94}
O.\ P.\ Sushkov, Phys.\ Rev.\ B49 (1994) 1250.

\bibitem{Dag94}
E.\ Dagotto, Rev.\ Mod.\ Phys.\ 66 (1994) 763.

\bibitem{Bel95}
V.\ I.\ Belinicher, A.\ L.\ Chernychev, A.\ V.\ Dotsenko, and O.\ P.\ Sushkov,
Phys.\ Rev.\ B51 (1994) 6076. 

\bibitem{Alt95}
J.\ Altmann, W.\ Brenig, A.\ P.\ Kampf, and E.\ M\"uller-Hartmann, Phys.\ Rev.\
B52 (1995) 7395.

\bibitem{Kyu97}
B.\ Kyung and S.\ I.\ Mukhin, Phys.\ Rev.\ B55 (1997) 3886.

\bibitem{Chu98}
A.\ V.\ Chubukov and D.\ K.\ Morr, Phys.\ Rev.\ B57 (1998) 5298.

\bibitem{Kar98}
N.\ Karchev, Phys.\ Rev.\ B57 (1998) 10913.

\bibitem{Ape98}
W.\ Apel, H.-U.\ Everts, and U.\ K\"orner, Eur.\ Phys.\ J.\ B5 (1998) 317.

\bibitem{Bru00}
M.\ Brunner, F.\ F.\ Assaad, and A.\ Muramatsu, Phys.\ Rev.\ B62 (2000) 15480.

\bibitem{Mis01}
A.\ S.\ Mishchenko, N.\ V.\ Prokof'ev, and B.\ V.\ Svistunov, 
cond-mat/0103234.

\bibitem{Sac03}
S.~Sachdev, Rev.\ Mod.\ Phys.\ 75 (2003) 913; Annals Phys.\ 303 (2003) 226.

\bibitem{Sus04}
O.\ P.\ Sushkov and V.\ N.\ Kotov, Phys.\ Rev. B70 (2004) 024503.

\bibitem{Kot04}
V.\ N.\ Kotov and O.\ P.\ Sushkov, Phys.\ Rev. B70 (2004) 195105.

\bibitem{Cha89}
S.~Chakravarty, B.~I.~Halperin, and D.~R.~Nelson, Phys.\ Rev.\ B39 (1989) 2344.

\bibitem{Wei79}
S.~Weinberg, Physica 96 A (1979) 327.

\bibitem{Gas85}
J.~Gasser and H.~Leutwyler, Nucl.\ Phys.\ B250 (1985) 465.

\bibitem{Kae05}
F.\ K\"ampfer, M.\ Moser, and U.-J.\ Wiese, Nucl.\ Phys.\ B729 (2005) 317.

\bibitem{Col69}
S.~Coleman, J.~Wess, and B.~Zumino, Phys.\ Rev.\ 177 (1969) 2239.

\bibitem{Cal69}
C.~G.~Callan, S.~Coleman, J.~Wess, and B.~Zumino, Phys.\ Rev.\ 177 (1969) 2247.

\bibitem{Geo84}
H.~Georgi, Weak Interactions and Modern Particle Theory, 
Ben\-ja\-min-Cum\-mings Publishing Company, 1984.

\bibitem{Gas88}
J.~Gasser, M.~E.~Sainio, and A.~Svarc, Nucl.\ Phys.\ B307 (1988) 779.

\bibitem{Jen91}
E.~Jenkins and A.~Manohar, Phys.\ Lett.\ B255 (1991) 558.

\bibitem{Ber92}
V.~Bernard, N.~Kaiser, J.~Kambor, and U.-G.~Meissner, Nucl.\ Phys.\ B388 (1992)
315.

\bibitem{Bec99}
T.~Becher and H.~Leutwyler, Eur.\ Phys.\ J.\ C9 (1999) 643.

\bibitem{Neu89}
H.~Neuberger and T.~Ziman, Phys.\ Rev.\ B39 (1989) 2608.

\bibitem{Fis89}
D.~S.~Fisher, Phys.\ Rev.\ B39 (1989) 11783.

\bibitem{Has90}
P.~Hasenfratz and H.~Leutwyler, Nucl.\ Phys.\ B343 (1990) 241.

\bibitem{Has91}
P.~Hasenfratz and F.~Niedermayer, Phys.\ Lett.\ B268 (1991) 231.

\bibitem{Has93}
P.~Hasenfratz and F.~Niedermayer, Z.\ Phys.\ B92 (1993) 91.

\bibitem{Chu94}
A.\ Chubukov, T.\ Senthil, and S.\ Sachdev, Phys.\ Rev.\ Lett.\ 72 (1994) 2089;
Nucl.\ Phys.\ B426 (1994) 601.

\bibitem{Leu94}
H.~Leutwyler, Phys.\ Rev.\ D49 (1994) 3033.

\bibitem{Hof99}
C.~P.~Hofmann, Phys.\ Rev.\ B60 (1999) 388; Phys.\ Rev.\ B60 (1999) 406; 
cond-mat/0106492; cond-mat/0202153.

\bibitem{Rom99}
J.\ M.\ Roman and J.\ Soto, Int.\ J.\ Mod.\ Phys.\ B13 (1999) 755; 
Ann.\ Phys.\ 273 (1999) 37; Phys.\ Rev.\ B59 (1999) 11418; 
Phys.\ Rev.\ B62 (2000) 3300.

\bibitem{Bae04}
O.~B\"ar, M.~Imboden, and U.-J.~Wiese, Nucl.\ Phys.\ B686 (2004) 347.

\bibitem{Wel95}
B.\ O.\ Wells, Z.-X.\ Shen, D.\ M.\ King, M.\ H.\ Kastner, M.\ Greven, and
R.\ J.\ Birgeneau, Phys.\ Rev.\ Lett.\ 74 (1995) 964.

\bibitem{LaR97}
S.\ La Rosa, I.\ Vobornik, F.\ Zwick, H.\ Berger, M.\ Grioni, G.\ Margaritondo,
R.\ J.\ Kelley, M.\ Onellion, and A.\ Chubukov, Phys.\ Rev.\ B56 (1997) R525.

\bibitem{Kim98}
C.\ Kim, P.\ J.\ White, Z.-X.\ Shen, T.\ Tohyama, Y.\ Shibata, S.\ Maekawa,
B.\ O.\ Wells, Y.\ J.\ Kim, R.\ J.\ Birgeneau, and M.\ A.\ Kastner, 
Phys.\ Rev.\ Lett.\ 80 (1998) 4245.

\bibitem{Ron98}
F.\ Ronning, C.\ Kim, D.\ L.\ Feng, D.\ S.\ Marshall, A.\ G.\ Loeser, L.\ L.\
Miller, J.\ N.\ Eckstein, I.\ Borovic, and Z.\ X.\ Shen, Science 282 (1998) 
2067.

\bibitem{Wei90}
S.~Weinberg, Phys.\ Lett.\ B251 (1990) 288; Nucl.\ Phys.\ B363 (1991) 3;
Phys.\ Lett. B295 (1992) 114.

\bibitem{Kap98}
D.~B.~Kaplan, M.~J.~Savage, and M.~B.~Wise, Phys.\ Lett.\ B424 (1998) 390; 
Nucl.\ Phys.\ B534 (1998) 329.

\bibitem{Epe98}
E.\ Epelbaum, W.\ Gl\"ockle, and U.-G.\ Meissner, Nucl.\ Phys.\ A637 (1998) 
107; Nucl.\ Phys.\ A684 (2001) 371; Nucl.\ Phys.\ A714 (2003) 535. 

\bibitem{Bed98}
P.~F.~Bedaque, H.-W.~Hammer, and U.~van~Kolck, Phys.\ Rev.\ C58 (1998) 641;
Phys.\ Rev.\ Lett. 82 (1999) 463; Nucl.\ Phys.\ A676 (2000) 357.

\bibitem{Kol99}
U.\ van Kolck, Prog.\ Part.\ Nucl.\ Phys.\ 43 (1999) 337.

\bibitem{Par99}
T.-S.\ Park, K.\ Kubodera, D.-P.\ Min, and M.\ Rho, Nucl.\ Phys.\ A646 (1999) 
83.

\bibitem{Epe01}
E.\ Epelbaum, H.\ Kamada, A.\ Nogga, H.\ Witali, W.\ Gl\"ockle, and U.-G.\ 
Meissner, Phys.\ Rev.\ Lett.\ 86 (2001) 4787.

\bibitem{Bea02}
S.\ Beane, P.\ F.\ Bedaque, M.\ J.\ Savage, and U.\ van Kolck, Nucl.\ Phys.\
A700 (2002) 377.

\bibitem{Bed02}
P.\ F.\ Bedaque and U.\ van Kolck, Ann.\ Rev.\ Nucl.\ Part.\ Sci.\ 52 (2002) 
339.

\bibitem{Nog05}
A.\ Nogga, R.\ G.\ E.\ Timmermans, and U.\ van Kolck, Phys.\ Rev.\ C72 (2005)
054006.

\bibitem{Bru05}
C.\ Br\"ugger, F.\ K\"ampfer, M.\ Pepe, and U.-J.\ Wiese, cond-mat/0511367.

\bibitem{Bru06}
C.\ Br\"ugger, C.\ Hofmann, F.\ K\"ampfer, M.\ Pepe, and U.-J.\ Wiese, in
preparation.

\bibitem{Zha90}
S.~Zhang, Phys.\ Rev.\ Lett.\ 65 (1990) 120.

\bibitem{Yan90}
C.~N.~Yang and S.~Zhang, Mod.\ Phys.\ Lett.\ B4 (1990) 759.

\bibitem{Ver00}
J.\ A.\ M.\ Vermaseren, ``New Features of FORM'', math-ph/0010025.

\bibitem{Abr72}
M.\ Abramowitz and I.\ A.\ Stegun, Handbook of Mathematical Functions,
Dover Publications, Inc., New York.

\bibitem{Lep97}
G.\ P.\ Lepage, nucl-th/9706029.

\bibitem{Mag66}
W.\ Magnus, ``Hills's equation'', Dover Publications, 1966.

\bibitem{Wie94}
U.-J.\ Wiese and H.-P.\ Ying, Z.\ Phys.\ B93 (1994) 147.

\bibitem{Bea96}
B.\ B.\ Beard and U.-J.\ Wiese, Phys.\ Rev.\ Lett.\ 77 (1996) 5130.

\bibitem{Cha03}
S.\ Chandrasekharan, M.\ Pepe, F.\ Steffen, and U.-J.\ Wiese, 
JHEP 0312 (2003) 035.

\end{thebibliography}
\end{document}